\documentclass[english]{article}
\usepackage[T1]{fontenc}
\usepackage[latin1]{inputenc}
\usepackage{amssymb}

\makeatletter

\newcommand{\lyxaddress}[1]{
\par {\raggedright #1
\vspace{1.4em}
\noindent\par}
}

\makeatletter

\date{}

\newcommand{\be}{\begin{equation}}
\newcommand{\ee}{\end{equation}}
\newcommand{\bea}{\begin{eqnarray}} 
\newcommand{\eea}{\end{eqnarray}} 

\newcommand{\ed}{\end{document}}
\newcommand {\rd} {{\rm d}}
\newcommand {\p} {\partial}

\makeatother

\usepackage{babel}
\makeatother

\begin{document}
\topmargin=0.0in

\title{
Piecewise constant  potentials and discrete ambiguities}

\author{ M. Lassaut and R.J. Lombard}

\maketitle

\lyxaddress{Institut 
de Physique Nucl\'{e}aire, IN2P3-CNRS, Universit\'e Paris-Sud 11, F-91406 
Orsay Cedex, France.}

\noindent
\begin{verbatim}
e-mail:  lassaut@ipno.in2p3.fr 
\end{verbatim}

\begin{abstract}

This work is devoted to the study of discrete ambiguities. For
parametrized potentials, they arise when the parameters are fitted to
a finite number of phase-shifts. It generates phase equivalent
potentials. Such equivalence was suggested to be due to the modulo
$\pi$ uncertainty inherent to phase determinations.
We show that a different class of  phase-equivalent potentials exists. 
To this aim, use is made of piecewise constant potentials, the intervals
of which are defined by the zeros of their regular solutions of the
Schr\"odinger equation.
 We give a classification of the ambiguities in terms of indices
 which include the difference
 between exact phase  modulo $\pi$ and the numbering
 of the wave function zeros.

\end{abstract}

pacs: 24.10.-i ; 03.65.-w ; 03.65.Nk

\section{Introduction}

Attempts to determine the potential from scattering data have a long
history, and conditions to obtain a unique answer are well known
(see for instance the textbooks by Newton \cite{newt}, 
Chadan and Sabatier \cite{CS}). 
 Approaches to the three-dimensional inverse scattering problem
can be classified in two categories \cite{newt,CS}.
In the first case, known  as  fixed-$\ell$ 
problem,  
the potential  
%
%
can be  constructed from the phase-shifts $\delta(\ell,k)$, if they are known  for all 
momenta $k\in (0,\infty)$, and from the discrete spectrum 
(bound state energies and the corresponding normalization constants). 
Note that the potential is  assumed to satisfy adequate integrability conditions \cite{CS}.
 In the second case, the so-called fixed-$E$ problem, Loeffel  \cite{Loef68} 
 has obtained theorems ensuring a unique 
potential from the knowledge of the phase-shifts  $\delta(\ell,k)$ 
for all (non-discrete) non negative values 
of $\lambda=\ell+1/2$. 
If the data set  reduces to discrete 
values of $\lambda=\ell+1/2$ for non-negative integer $\ell$, the 
Carlson's theorem \cite{Carlson}  predicts a unique potential   $V(r)$, provided 
it belongs to a suitable class \cite{CS,Loef68}. 

The present work is dealing with the second aspect of the problem.
It is relevant to the case where a chosen 
parametric form is used to fit a differential cross-section at a fixed energy.
Here, the uniqueness is
defined in the sense of a best fit to a differential cross section,
for instance. 
The argument does not apply to
  phase shifts, which can be 
 extracted from the scattering amplitude, since a phase is defined $mod \ \  \ n\pi$. Such
undeterminations are called phase- or 
discrete ambiguities. They have been
noticed long ago by Drisko, Satchler and Bassel \cite{DSB}.

As a concrete example, let us quote the discrete ambiguities described
 by Goldberg {\it et al} \cite{Goldberg}.  By using potentials of Woods-Saxon shape,
these authors have extracted 
four $\alpha-^{208} {\rm Pb}$ 
potentials, characterized by depth differences of 50 MeV in the real potential part, 
giving similar $\chi$-square per degree of freedom. 
The $\chi's$ square are related to the 
phase-shifts via the fit to the differential cross sections.
The   Jeffreys, Wentzel, Kramers and Brillouin (JWKB) phase-shifts have been calculated for the real parts
of the four potentials.  They  were found to differ by a factor $n \pi$, 
$n$ integer and independent of $\ell$
for low enough angular momentum ($\ell \leq 44$).

Attempts to study these ambiguities are scarce.
Sabatier \cite{Sabatier} and Cuer \cite{Cuer} have studied their 
origin by means of the JWKB approximation. 
A physical interpretation has been given by Leeb and Schmid \cite{LB}, 
in which the occurrence of 
discrete ambiguities is linked to the existence of partly Pauli forbidden states.

 Note that  the discrete ambiguities are still of interest.
 Let us mention the recent papers on the subject by Brandan \cite{MEB95} 
and by Amos and  Bennett \cite{AB97}. These works have been done in the context of  
  heavy and light nuclear ion scattering,  respectively.

The purpose of the present work is a further  study of the
discrete ambiguities.
In order to discuss this problem in  detail, and stress the
origin of the modulo $\pi$ shift, we shall consider piecewise-constant
 real potentials. Such potentials have been used recently by
Ramm and Gutman \cite{RG}. These authors  have shown that different
piecewise-constant  
positive real potentials can lead to almost the same 
fixed-energy phase-shifts. 
As we shall see, this is a convenient
starting point for the present study.  
Here we enlarge the study and consider also negative potentials.
Moreover, we found it particularly useful to restrict the class of 
potentials to the subset of piecewise-constant potentials the
intervals of which are defined by the zeros of 
the regular solution of the Schr\"odinger
equation.

The present work shows that piecewise constant potentials  are a basis for a
class of phase equivalent potentials, with a phase ambiguity of $n
\pi$. Within each $n$, 
these potentials can be ordered according to the positions of the zeros
of the regular solutions. The value of $n$ has a minimum, which 
 can be calculated by means of the JWKB approximation.

 Dealing with the exact solution of the Schr\"odinger equation, our work 
goes beyond, for a certain class of potentials, the earlier
 attempts by Sabatier \cite{Sabatier} and Cuer \cite{Cuer},
 who have resorted  to the JWKB approximation. 
The drawback is that our method becomes very tedious as the number of partial waves increases.

In what concerns numerical applications,
 the energy  will be put equal to unity, without loss of generality.  
 Indeed, use is made of the following scaling property:
 if $\delta(\ell,k=1)$ is the phase for the potential $V(r)$ at the energy $E=1$,
 it is also the phase for the potential $k^2 V(kr)$ at the energy $E=k^2$.

The paper is organized as follows. In section {\bf 2} the basic formalism is
recalled. In section {\bf 3} our results are presented.
The results are discussed in the JWKB approximation in section {\bf 4}  and our conclusions are presented 
in section {\bf 5}.

\section{Formalism}

In this paper we investigate to what extent  piecewise constant 
potentials  allow us to reproduce 
 a finite set of fixed-energy phase-shifts. 
Considering spherically symmetric potentials of finite range,
we start from the reduced radial 
Schr\"odinger equation. For a positive energy $E=k^2$ and an    
 angular momentum $\ell$ ($\hbar = 2m = 1$), it reads 
\be
      \psi_{\ell}(k,r)'' + \left(k^2- v(r) -\frac{\ell (\ell+1)}{r^2} \right) 
      \psi_{\ell}(k,r) = 0 \ .
\label{Schr1}
\ee
Here, the reduced potential is defined by 
\be
v(r)=\frac{2 m}{\hbar^2} V(r) \ ,
\label{vred}
\ee
in terms of the potential $V(r)$, and
$\psi_{\ell}(k,r)$ is the regular solution of Eq.(\ref{Schr1}) defined 
 by the Cauchy condition $\lim_{r \to 0} \psi_{\ell}(k,r) r^{-\ell-1}=1$.

 Our construction of phase-equivalent potentials is based upon the 
 following property.
Suppose the function  $\psi_{\ell}(k,r)$ to vanish 
for $r=r(\ell,k)$. 
It is unnecessary to know the potential for
$ r < r(\ell,k)$ to determine the phase shift $\delta(\ell,k)$. 
This holds for $(\ell,k)$ fixed.
Thus, it
is sufficient to consider the Schr\"odinger equation on the interval 
$[r(\ell,k),\infty[$. Actually, the derivative $\psi'_\ell$ with
respect to $r$ is undetermined at $r = r(\ell,k)$, 
which has no consequence since the
phase-shift is determined only by
the ratio $\psi_{\ell}/\psi'_{\ell}$.
Furthermore, for a fixed energy 
and a fixed $\ell$, there is a countable number of zeros \cite{LLSW}.
The functions  $\psi_{\ell}(k,r)$ and 
$\psi_{\ell}'(k,r)$  cannot vanish at the same value of $r$ for $2 \ell+1 >0$ 
(except
 for $r=0$) and $\ell >0$ as shown in \cite{Hille}. 
 Consequently, $\psi_{\ell}(k,r)$ 
has only simple zeros (except at $r=0$). 
We recall that thanks to the fact that the zeros are simple,
 they satisfy the strict ordering 
 $ 0 < \hat r_1(\ell,k) <  \hat r_2(\ell,k) < 
 \dots < \hat  r_n(\ell,k) < \ldots < R$.

The $n$th zeros of the regular solution  
satisfy a monotony property, namely, for fixed $n$, the function   
\be
\ell \mapsto \hat r_n(\ell,k)
\ee
increases with $\ell$ as has shown by Sturm in the 1830s \cite{Sturm}.
More precisely we have

\be
      \frac {\rd}{\rd \ell} \hat r_n(\ell,k) =
     (2 \ell+1) \frac{\int_{0}^{\hat r_n(\ell,k)} \rd r' \ \psi^2(k,r')}{
      \left (\displaystyle \frac{\p}{\p r} 
          \psi_\ell(k,\hat r_n(\ell,k)) \right)^{2}}  \ ,
\label{zerol}
\ee
which is positive definite. As stated before, 
$\psi_{\ell}$ and $\psi_{\ell}'$  cannot vanish
 simultaneously for  $2 \ell+1 >0$ , except for $r=0$ and $\ell >0$
 \cite{Hille}, so that the denominator is never equal to zero. 

To find potentials reproducing  $N+1$ fixed energy $E=k^2$ 
phase-shifts, 
$\delta(\ell_j,k), \ell=0,1, \dots N$, we apply the following procedure :

\begin{enumerate}
\item
 We construct a piecewise constant potential with a compact support: $v(r)=0$ for $r \geq R$,
$R$ being fixed.

\item
$\hat r_n(\ell,k)$ is the n-th zero  of the regular solution  
$\psi_{\ell}(k,r)$
 of the Schr\"odinger equation (\ref{Schr1})  for the $\ell$-wave  aside from the
trivial zero at the origin.  
We denote by $r_{\ell}$ a generic non trivial zero of    $\psi_{\ell}(k,r).$
 For the zero potential  $\hat r_n(\ell,k)$ is given in terms of the 
 zeros of the spherical Bessel function 
 $j_{\ell}(z)$  labelled $x_{\ell,n}$. We have  
 $k \hat r_n(\ell,k) \equiv x_{\ell,n}$.

\item
The potential  $v$ of Eq.(\ref{vred}) 
is assumed to be constant on each interval delimited by the partition 
$ 0 < r_1 < r_2 < r_3 , \ldots < r_N < R$. We set
\bea
v(r) &=& v_0,\ \  K_0^2 =k^2-v_0 \ \ \  ;  \ \ \ 0 \leq r  < r_1  \nonumber\\ 
v(r) &=& v_1,\ \  K_1^2 =k^2-v_1 \ \ \ ;  \ \ \ r_1  \leq r  < r_2  \nonumber\\ 
 & & \dots   \nonumber\\ 
 v(r) & = & v_N,\ \ K_N^2=k^2-v_N \ \ \ ;\ \ \  r_N \leq  r <  R
\eea

\item 
The regular $u_{\ell}(k r)$ and irregular $w_{\ell}(k r)$ 
solutions of  Eq. (\ref{Schr1}) for $v \equiv 0$ are denoted, respectively, 
\bea 
u_{\ell}(x) & = & \sqrt{\frac{\pi x}{2}} J_{\ell+1/2}(x) \nonumber\\
w_{\ell}(x) & = & -  \sqrt{\frac{\pi x}{2}} Y_{\ell+1/2}(x) \nonumber
\eea
in terms of the    Bessel    functions $J_{\nu},Y_{\nu}$ of order $\nu$, 
given in \cite{Erd}.
We have $u_{\ell}(x)=x j_{\ell}(x)$ where $j_{\ell}$ is the spherical Bessel function of order $\ell$.

\item
Let  $\psi_{\ell}$ be the regular solution of the Schr\"odinger 
equation (\ref{Schr1}). 
 We denote by $A(r),B(r),C(r),D(r),...$ the value of the ratio 
\be
\frac{\psi_{\ell}(k,r)}{\psi'_{\ell}(k,r)} \ 
\label{ratioE}
\ee
for $\ell=0,1,2,3,...$ respectively. 
Here, the prime denotes the first derivative with respect to $r$.

\end{enumerate}

As we shall see, these requirements give rise to a class of  
phase-equivalent potentials.

To show phase-ambiguities,
it is worth to recall that the exact phase can
be calculated by using the variable phase
method of Calogero \cite{Calo}. Most of the time it allows the removal of  
the $n \pi$ ambiguity burdening the direct solution.
With this method, the phase-shift is reached
by solving a first order differential equation 
\be
\frac{\partial}{\partial r} \delta(\ell,k,r) =-\frac{v(r)}{k} 
\ (u_{\ell}(k r)  \cos(\delta(\ell,k,r))+  w_{\ell}(k r)  
\sin(\delta(\ell,k,r)) )^2 \ ,
\label{cal}
\ee
with $\delta(\ell,k,0) = 0$ as boundary condition. The phase-shift 
is  given by the limit $\delta(\ell,k) =\lim_{r \to \infty} \delta(\ell,k,r)$.
 In the present work, use is made of the variable phase method to determine the exact
phases.  In other words, $k$ being fixed, we solve Eq.(\ref{cal}) for
 each value of $r$ starting from $r=0$, combined with $\delta(\ell,k,0)=0$.  
Actually, $\lim_{r \to \infty} \delta(\ell,k,r) =\delta(\ell,k,R) \equiv \delta(\ell,k).$ 

 A variant of the Calogero's equation 
 but for  $\tan( \delta(\ell,k,r))$ has been derived in \cite{HLW}.
 It can be obtained from Eq.(\ref{cal}) noticing that 
\be
\frac{\partial}{\partial r} \tan(\delta(\ell,k,r))=\frac{1}{\cos(\delta(\ell,k,r))^2} \ 
 \frac{\partial}{\partial r} \delta(\ell,k,r) \ .
\ee
Setting $T(\ell,k,r)= \tan( \delta(\ell,k,r))$, it yields
\be
\frac{\partial}{\partial r} T(\ell,k,r) =-\frac{v(r)}{k} 
\ (u_{\ell}(k r)  +  w_{\ell}(k r)  
T(\delta(\ell,k,r)) )^2 \ ,
\label{calt}
\ee
which is solved with the boundary condition $T(\ell,k,0)=0$.
At zero energy, the latter formalism  allows  the calculation of 
both the scattering length and the effective range directly from the potential.
In our case, we have chosen to work with Eq.(\ref{cal}) to avoid the numerical difficulties owing to  poles in $T(\ell,k,r)$.

\section{Examples}

\subsection{A single phase-shift}

Although the case of a single known phase-shift is trivial,  
it illustrates features of the discrete ambiguities.

Assume, for example, the s-wave phase-shift 
$\delta(\ell=0,k) \equiv \delta_0$ to be known.
The procedure amounts  to find a constant $v_0$ in $[0,R]$. 
Setting 
\be
\frac{\tan(k R + \delta_0)}{k}  = A(R) \ , \ee
the equation to be solved is, depending upon whether 
 $K_0^2$ is positive or negative
\bea
\frac{\tan(K_0 R)}{K_0}& = & A(R) \ , \ \ \ K_0 = \sqrt{K_0^2}  \nonumber\\
 \frac{\tanh( K_0^*   R)}{ K_0^* } & = & A(R) \ , \ \ \ K_0^* =
 \sqrt{-K_0^2}
\label{basic0} 
 \ . \eea

For $K_0^2 \geq 0$, discrete ambiguities come from the fact there is 
a countably infinite set of solutions to Eq.(\ref{basic0}). 
Whatever the right-hand-side of (\ref{basic0}), 
one solution of (\ref{basic0}) exists in the interval
\be
(2 n -1) \frac{\pi}{2} \leq \sqrt{k^2-v_0} \ R < (2 n +1) \frac{\pi}{2}  
\label{amb1d}
\ee
for every $n \geq 1$. In $[0,\pi/2[$ the solution exists only for 
suitable values of the r.h.s. of the equation.
The difference between the solutions for $v_0$ on two adjacent
intervals  is of the 
order of the square of the length of the interval, namely 
$\pi^2 $. At the potential level, it introduces a difference of 
$\Delta V \sim \frac{\pi^2 \hbar^2}{2 m R^2}$. As a consequence,
the larger  is the mass  $m$, the smaller are the differences between phase
equivalent potentials, 
in agreement what is observed phenomenologically \cite{Mich}. It means
that while searching optical potential parameters, the chance to find
phase equivalent potentials on a finite parameter space is larger for
heavy than for light systems.  
Indeed discrete ambiguities are often observed for 
nucleus-nucleus scattering with a heavy target, for instance for  
$\alpha-^{208} {\rm Pb}$ scattering (not so for  $\alpha-^{58} {\rm Ni}$),  whereas few 
attempts have been made for nucleon-nucleus scattering.
Note that for  proton scattering on Cr  isotopes  at 10 MeV, 
discrete ambiguities have been reported by Andrews {\it et al.} \cite{And}. 
The difference in the potential depth is about 100 per cent.
At high energies $\vert v_0 \vert$ is weak as compared to $ k^2$,  and we expect 
$\sqrt{k^2 -v_0} R$ to fall in the same interval  
as $k R$, namely 
\be
(2 n_k -1) \frac{\pi}{2} \leq k  \ R < (2 n_k +1) \frac{\pi}{2} \ .  
\ee
This is due to the fact that $\sqrt{k^2-v_0} R \sim k R -v_0 R/(2 k)$. As
a consequence, the solution cannot be found in another interval, and the
discrete ambiguities disappear. This is confirmed by the analysis of
scattering data at high energies.
Finally, to finish with this example, 
we have verified that the different solutions of Eq. (\ref{amb1d}) lead to s-wave phase 
 shifts which differ by multiples of $\pi$.

\subsection{2 phase-shifts $\ell=0,1$}

The determination of the piecewise constant potential starts from the
largest zero $r_N < R$ fitting the solution  $v_N (K_N)$ on the largest 
$\ell=N$ phase-shift considered. 
Finding $v_{N-1} (K_{N-1})$ on the previous interval 
$[r_{N-1},r_N[$ is more complicated in
general, as it depends not only on $r_{N-1}$ but also from 
$r_{N},K_N$.  The complexity increases with the number of
zeros considered. 
Two exceptions exist, namely the $\ell=0,1$  and $\ell=0,1,2$ cases.  
These exceptions are called simply soluble, and their solutions are presented below.

The regular solution for $\ell=1$ is denoted 
 $u_1(K_0 r)$ for  $r < r_1$. Since $r_1$ is a zero  of $\psi_1$,  we have     
  $K_0 r_1=x_{1,n_1}$,
and
\be
(\forall n_1) \ u_1(x_{1,n_1}) =\sin(x_{1,n_1})/x_{1,n_1}-\cos(x_{1,n_1}) 
\equiv 0 \ .
\label{eqzer1}
\ee
On  the interval $[r_1,R[$, up to a constant factor, it  
reads :
\be
\psi_1(k,r)=u_1(K_1 r) w_1(K_1 r_1)-u_1(K_1 r_1) w_1(K_1 r) \ .
\label{ex1}
\ee
Finally, for $r \geq R$ the regular solution is proportional to 
\be
\psi_1(k,r)=u_1(k r) \cos(\delta_1) +w_1(k r) \sin(\delta_1) \ .
\label{ex2}
\ee
Equating the inverse of the  logarithmic derivative of solutions 
(\ref{ex1},\ref{ex2}) at $r=R$, we obtain
\bea
\frac{ \tan(K_1 (R-r_1)) }{K_1}&= &
\frac{\mathcal{N}}{\mathcal{D}} 
\qquad\quad  K_1=\sqrt{  K_1^2}  \nonumber\\
\frac{ \tanh(K_1^* (R-r_1)) }{K_1^*} & =&
\frac{\mathcal{N}}{\mathcal{D}} \qquad\quad K_1^*=\sqrt{- K_1^2}    \ \ .
\label{eql1}
\eea
Here we have
\bea
 \mathcal{N} & = & -B(R) R - R^2 + B(R) r_1 + R r_1 - B(R) K_1^2 R^2 r_1 \nonumber\\
 \mathcal{D} & = & -B(R) - R + B(R) K_1^2 R^2 - B(R) K_1^2 R r_1 - K_1^2 R^2 r_1 \ .
\label{eql1p}
\eea

We recall that in these expressions, $B(R)=(\psi_1/\psi'_1)(k,R)$, 
 with $\psi_1$ given by  Eq.(\ref{ex2}), at $r \geq R$.

Turning to the $\ell=0$ wave, the regular solution in $[0,r_1[$ 
is proportional to  $\sin(K_0 r)$. Therefore,   
$(\psi_0/\psi'_0)(r=r_1)=\tan(K_0 r_1)/K_0$, which is equal to $r_1$ 
when Eq. 
 (\ref{eqzer1}) is taken into account. At the other end of
the interval, $[r_1,R[$, the function $\psi_0/\psi'_0$ is given by
\be
\frac{\psi_0(k,R)}{\psi_0'(k,R)}=\frac{K_1r_1\ 
\cos( K_1(R - r_1)) + \sin( K_1(R - r_1))}{ K_1 
(\cos( K_1(R - r_1)) - K_1r_1\ \sin( K_1(R - r_1)))} \ .
\ee
It is identified  with $A(R)=\tan(k R + \delta_0)/k$.  We then obtain
\bea
\frac{\tan(K_1 (R-r_1))}{K_1}&=&\frac{A(R) - r_1}{1 + A(R)  \ K_1^2 \  r_1} 
 \qquad\quad K_1=\sqrt{K_1^2}  \nonumber\\
\frac{\tanh(K_1^* (R-r_1))}{K_1^*}&=&\frac{A(R) - r_1}{1 + A(R)  \ K_1^2 \  r_1} 
 \qquad\quad K_1^*=\sqrt{-K_1^2} \ .
\label{eql01}
\eea
Equating equations (\ref{eql1}), (\ref{eql1p}) and( \ref{eql01}),  
we are left with 
\be
K_1^2=\frac{A(R) B(R) + A(R) R - B(R) R - R^2}{A(R) B(R)  R^2} \ .
\label{solk1}
\ee
Provided that $A(R) B(R) \ne 0$ or equivalently that $R$ is  a zero
 of neither $\psi_0$ nor $\psi_1$.  The equation (\ref{solk1}) also reads
\bea
K_1^2&=&k^2+\frac{{\cal N}}{{\cal D}} \nonumber\\
{\cal N}&=&k^3 R \sin(\delta_0 - \delta_1) \nonumber\\ 
{\cal D} & = & \sin(k R + \delta_0) \ (\sin(k R + \delta_1)-  
k R \cos(k R + \delta_1))  \ .
\eea
It fixes the value of $K_1^2$,  and consequently the constant $v_1$, 
on the interval $[r_1,R[$ independently  of the value of $r_1$. 
For this very reason, the model
 is quoted as simply soluble.
With this value of $K_1^2$,  Eq. (\ref{eql01}) 
allows us to determine $r_1$. 
In this case, a first source of ambiguities come from 
the choice of the value of $r_1$ among   
the zeros of $\psi_1$ smaller than $R$. The next step consists
in determining $v_0$. To this aim, use is made of the fact that
$K_0 r_1 =x_{1,n_1}$ is a zero of   $\psi_1$  $\forall n_1$.
The choice of  which zero $x_{1,n_1}$ of the spherical Bessel  function $j_1$ is another source of ambiguity.

{\bf Numerical application}

To fix ideas, the following example is treated. A reference potential
is defined by setting
$v_0=-3, \ r \in [0,r_1[$,  $v_1=-8, \ r \in [r_1,R[$. The energy is 
fixed at $E=k^2=1$.   Thus, we have 
$K_0=2$ and $K_1=3$.  
The first zero of the spherical Bessel function $j_1$  is chosen,
namely $x_{1,1}=4.493409$.
It fixes the value of $r_1$ at $2.2467045$.
 The two phase-shifts $\delta_0$
and $\delta_1$ are then calculated by the variable phase method.   

With these pseudo-data,
we look for  phase-equivalent potentials of the same range $R=10$, i.e. 
potentials giving the same $\ell=0,1$-phase-shifts modulo $\pi$.
In fact the phase-equivalent potentials are determined from the 
values of $A(R),B(R),R=10$ 
calculated from the reference potential. We recall that $K_1^2$ 
is recovered from Eq.(\ref{solk1}), and is 
 therefore identical to the starting value.  
In the Table  1 are shown the different values of $r_1 < R=10$ 
obtained  from Eqs.(\ref{eql01}). 
Only the    first two zeros of the spherical Bessel function $j_1(z)=u_1(z)/z$ 
namely $x_{1,1}=4.493409, x_{1,2}=7.725252 $, are  displayed  in  
Table  1. They  concern  the ambiguities relating to $K_0$.

\vspace*{.2cm}
\begin{table}
\begin{center}
\begin{tabular}{|c|c|c|c|c|}
\hline
\multicolumn{2}{|c}{ $x_{1,1}= 4.493409$} & \multicolumn{2}{|c|} 
 {$x_{1,2}=7.725252  $}   &   \\
\hline
$n$ & $K_0$  & $n $ & $K_0$ & $r_1 $ \\
\hline
-7 &  0.467351105& -6&  0.803489067&  9.61463257 \\
 -6&  0.524561926& -5&  0.901848241&  8.5660223 \\
 -5&  0.597764867& -4&  1.02770172&  7.51701782 \\
 -4&  0.694775396& -3&  1.19448618&  6.467427 \\
 -3&  0.829515446& -2&  1.42613676&  5.41690836 \\
 -2&  1.02946743& -1&  1.76990236&  4.36479003 \\
 -1&  1.35771439& 0&  2.33423792&  3.30953931 \\
 0&  1.99999964& 1&  3.4384809&  2.24670498 \\
 1&  3.89058288& 2&  6.68884865&  1.15494498 \\
\hline
\end{tabular}
\end{center} 
\caption{  For the first two zeros of $j_1(x)$, the value of $K_0$
is listed as function of the shift ($n \pi$) with respect to the exact
phase of the reference potential $K_0=2, r_1= 2.24670498, K_1=3,R=10$.}
\end{table}

The ambiguities proceed 
from two sources: the different values of $r_1$ and the choice of 
the zero $x_{1,n_1}$. As a consequence,  
two different potentials may give  
the same shift with respect to the exact phase. 
At this stage, the ambiguities can be characterized by two numbers: \\
1)  $n_{1}$ : numbering the zero of the regular solution 
for the p-wave ($\ell=1$) \\
2) $n$ : the difference between the exact phases of
 the phase-equivalent  and the starting potentials
divided by $\pi$. In the
present example this number happens to be the same for  both   $\ell=0$  
and  $\ell=1$ phases.

The values of $K_0$ are listed and ordered as function of the $n \pi$
shift with respect to the exact solution. For $n=0$  and 
$x_{1,1} = 4.493409$, the parameters of the reference potential are recovered with
a $10^{-7}$ accuracy. Due to the finite range $R$,  Table  1  exhausts
the number of phase equivalent potentials for the first two zeros of
$j_1$. Other cases exist, corresponding to the higher zeros of $j_1$.
Here $n_1=1$ for all $K_0$ of the second column and  
$n_1=2$ for the $K_0$'s of column 4.

It is interesting to quote results for 
 transparent potentials. They are characterized by  $\delta_0=\delta_1=0$. 
It implies  automatically $K_1=k=\sqrt{E}=1$ and $r_1$ takes only the two values    
$7.72525183$ or   $4.49340947$ smaller than the range parameter $R = 10$.
A freedom still exists in the choice of $x_{1,n_1}$ which determines $K_0$.
The results are displayed in  Table 2.

\begin{table}
\begin{center}
\begin{tabular}{|c|c|c|c|c|}
\hline
  $n $ & $n_{1}$ & $K_0$ & $r_1 $  \\
\hline
  -1& 1 & 0.581652127&  7.72525183\\
  0&  1 &0.999999931&  4.49340947\\
  0&2 &   1.00000004&  7.72525183\\
  1& 2 & 1.71924064&  4.49340947\\
  1& 3 & 1.41149086&  7.72525183\\
  2& 3 & 2.42669234&  4.49340947\\
  2& 4 &  1.82080713&  7.72525183\\
  3&  4 &3.13040547&  4.49340947\\
  3& 5 & 2.22915123&  7.72525183\\
  4& 5 &  3.83244722&  4.49340947\\
  4& 6 & 2.63697599&  7.72525183\\
  5& 6 & 4.53359608&  4.49340947\\
  5& 7 & 3.0444899&  7.72525183\\
  6& 7 & 5.23421053&  4.49340947\\
\hline
\end{tabular}
\end{center}
\caption{Transparent potentials $R=10$. For two values of
$r_1=x_{1,1},x_{1,2}$, $K_0$ is listed as a function of  $n $ defined in the text,
together with $n_{1}$, the order of the zero of $j_1(x)$.}
\end{table}

\subsection{3 phase-shifts $\ell=0,1,2$}

Consider the case where three phase-shifts 
$\delta_0,\delta_1,\delta_2,$ corresponding 
to $\ell=0,1$ and $2,$ are known.
On the interval $[r_2,R[$, the $\ell =2$ regular solution  reads, 
  up to a constant multiplicative factor,
\be
\psi_2(k,r)=u_2(K_2 r) w_2(K_2 r_2)-u_2(K_2 r_2) w_2(K_2 r) \ .
\ee
It vanishes at $r=r_2$ as imposed. For $r \geq R$, it is proportional to 
\be
\psi_2(k,r)=u_2(k r) \cos(\delta_2) +w_2(k r) \sin(\delta_2) \ .
\label{psi2}
\ee
With  $C(R)=(\psi_2/\psi'_2)(k,R)$, the continuity conditions at $r=R$
require

\bea
\frac{\tan(K_2 (R-r_2))}{K_2}&  = &\frac{\mathcal{N}}{\mathcal{D}}\ , 
\ \ \ \ K_2=\sqrt{K_2^2}  \nonumber\\
\frac{\tanh(K_2^* (R-r_2))}{K_2^*} &  =& \frac{\mathcal{N}}{\mathcal{D}}\ , 
\ \ \ \ K_2^*=\sqrt{-K_2^2}
\ . 
\label{T22mp}
\eea
Here we have 
\bea
\mathcal{N} & = & -18  C(R) R - 9  R^2 + 3  C(R)  K_2^2  R^3 + 18  C(R)   r_2 + 9    R  r_2 - 
     9  C(R)  K_2^2  R^2  r_2 \nonumber\\
  & -& 3  K_2^2  R^3  r_2 + 6  C(R)  K_2^2  R  r_2^2 + 3  K_2^2  R^2  r_2^2 - 
     C(R)  K_2^4  R^3  r_2^2 \nonumber\\
\mathcal{D} &= &-18  C(R) - 9  R + 9  C(R)  K_2^2  R^2 + 3  K_2^2  R^3 - 
     18  C(R)  K_2^2  R  r_2 - 9  K_2^2  R^2  r_2 \nonumber\\
 &  +& 3  C(R)  K_2^4  R^3  r_2  + 6  C(R)  K_2^2  r_2^2 + 
     3  K_2^2  R  r_2^2 - 3  C(R)  K_2^4  R^2  r_2^2 \nonumber\\ 
  &-& K_2^4  R^3  r_2^2 \ .
\label{T22mpp}
\eea

Other constraints come from the fact that $r_2$ is a zero of $\psi_2$.
First, for $0 \leq r <  r_1$, the function $\psi_2$ is proportional 
to $u_2$. 
At $r=r_1$, we have $(u_2/u'_2)(K_0 r_1)=-r_1/2$, 
since for $x=K_0 r_1$  we have $\sin(x)/x-\cos(x)=u_1(x)=0$.
Expanding $\psi_2$ as a linear combination of $u_2(K_1 r)$ and 
$w_2(K_1 r)$ (in the case where $K_1^2 \geq 0, K_1=\sqrt{K_1^2}$) 
on $[r_1,r_2[$, taking into account that 
$(\psi_2/\psi'_2)(k,r)\  =\  -r_1/2$ at 
 $r=r_1$ and  zero at $r=r_2$, we are left with:
\bea
\frac{\tan(K_1 (r_2-r_1))}{K_1}&=& \frac{\mathcal{N}}{\mathcal{D}} 
\qquad\quad   K_1=\sqrt{ K_1^2}  \nonumber\\
\frac{\tanh(K_1^* (r_2-r_1))}{K_1^*}&=& \frac{\mathcal{N}}{\mathcal{D}} 
\qquad\quad   K_1^*=\sqrt{- K_1^2}  \nonumber\\
 \frac{\mathcal{N}}{\mathcal{D}} &= & 
 \frac{-3  r_1 + 3  r_2 +  K_1^2  r_1  r_2^2 }{3 + 3  K_1^2  r_1  
 r_2 -  K_1^2  r_2^2}   \ .
\label{eqr1}
\eea

For $\ell=1$, the regular solution  is zero at $r=r_1$, and reads 
at $r=r_2$, up to a non zero multiplicative factor, 
(when $ K_1^2 \geq 0,   K_1=\sqrt{K_1^2} $)
\be
\psi_1(k,r_2)  =  u_1(K_1 r_2) w_1(K_1 r_1)-  w_1(K_1 r_2) u_1(K_1 r_1),
\ee
and its derivative is
\be
\psi'_1(k,r_2)=K_1 (  u_1'(K_1 r_2) w_1(K_1 r_1)-  w_1'(K_1 r_2) u_1(K_1 r_1)) \ .
\ee
It implies that
\be
\frac{r_2 \ (r_1 + \mathcal{T} + r_2 \ (-1 + K_1^2 \ r_1 \ \mathcal{T}))}
{-r_1 + r_2 - \mathcal{T} - K_1^2 \ r_1 \ r_2 \ \mathcal{T} + 
  K_2^2 \ r_2^2 \ (r_1 + \mathcal{T})}= 
  \frac{\psi_1(k,r_2)}{\psi'_1(k,r_2) } \ . 
\label{eql1r2}
\ee
Here we have defined
\bea
\mathcal{T} &=&\frac{\tan(K_1 (r_2-r_1))}{K_1} 
\qquad\quad \ K_1=\sqrt{K_1^2}  \nonumber\\
\mathcal{T} &= & \frac{\tanh(K_1^* (r_2-r_1))}{K_1^*} 
\qquad\quad \ K_1^*=\sqrt{-K_1^2} \ .
\label{eql1r2p}
\eea
Equating  $\mathcal{T}$ with $\mathcal{N}/\mathcal{D}$ of Eq.(\ref{eqr1})  
and using  Eq.(\ref{eql1r2}),  we obtain $(\psi_1/\psi'_1)(k,r_2)=B(r_2)=r_2/2$.
 
The $\ell =1$ regular solution $\psi_1$ on  $[r_2,R[$ is a  linear combination of 
$u_1(K_2 r)$ and $w_1(K_2 r)$ 
(when $ \ K_2^2 \geq 0,    K_1=\sqrt{K_1^2}  $). 
Using  the constraint that $(\psi_1/\psi'_1)(k,r)$ is equal to $r_2/2$ at $r=r_2$ and to $B(R)$ at $r=R$ we obtain: 

\bea
\frac{\tan(K_2 (R-r_2))}{K_2}& =& \frac{\mathcal{N}}{\mathcal{D}} 
\qquad\quad  \ K_2=\sqrt{ K_2^2}  \nonumber\\
\frac{\tanh(K_2^* (R-r_2))}{K_2^*}& =& \frac{\mathcal{N}}{\mathcal{D}} 
\qquad\quad K_2^*=\sqrt{-K_2^2} \label{T22p0} \\
\mathcal{N} & = & 3  B(R) R + 3 R^2 - 3  B(R)  r_1 - 3 R  r_2 \nonumber\\
  &+ & 3  B(R) K_2^2 R^2  r_2 -  
  B(R) K_2^2 R  r_2^2 -   K_2^2 R^2  r_2^2 \nonumber\\
\mathcal{D} & = & 3  B(R) + 3 R - 3  
B(R) K_2^2 R^2 + 3  B(R) K_2^2 R  r_2  \nonumber\\
&+ &  3 K_2^2 R^2  r_2   -   B(R) K_2^2  r_2^2 - K_2^2 R r_2^2 \nonumber\\
 & + &  B(R) K_2^4 R^2  r_2^2 \ .
\label{T22p}
\eea

Identifying (\ref{T22mp}, \ref{T22mpp}) and (\ref{T22p0},\ref{T22p}) we have finally

\be
K_2^2=\frac{4  B(R)  C(R) + 2  B(R) R - 2  C(R) R - R^2}{ B(R)  C(R) R^2}
\label{eqk2}
\ee 
or, equivalently, provided that $R$ is not a zero of  $\psi_1$ or $\psi_2$
\bea
K_2^2&=&k^2+\frac{{\mathcal N}}{{\mathcal D}} \nonumber\\
{\mathcal N}&=&k^5 R^3 \sin(\delta_1 - \delta_2) \nonumber\\ 
{\mathcal D} & = &(\sin(k R + \delta_1)-  k R \cos(k R + \delta_1))  \nonumber\\
              & \times &  ((3 - k^2 R^2) \sin(k R + \delta_2) -
                     3 k R \cos(k R + \delta_2))  \ .
\label{eqk2p}
\eea
Here again the value of $K_2^2$ and of $v_2$ on $[r_2,R[$ 
does not depend on $r_2$, and
 the model is still simply soluble. 
Once $K_2^2$ is known,  we have to determine $r_2$ 
through Eqs. (\ref{T22p0},\ref{T22p}).  
Then Eq. (\ref{solk1}) expresses $K_1^2$ in terms of the ratio 
$(\psi_{\ell}/\psi'_{\ell})(k,r_2)$ for $\ell=0,1$, namely the quantities 
$A(r_2),B(r_2)$.
We recall that $(\psi_{1}/\psi'_{1})(k,r_2)=r_2/2 = B(r_2)$, 
as shown before. 
By definition  $A(r_2) =(\psi_{0}/\psi'_{0})(k,r_2)$. Thus, 
on the interval $[r_2,R[$, the matricial equation for $K_2^2 \geq 0$
\be
 \left( \matrix{\cos(K_2 (R-r_2))  &  \frac{\sin(K_2 (R-r_2))}{K_2} \cr
-K_2 \ \sin(K_2 (R-r_2)) &\cos(K_2 (R-r_2))   \cr} \right) 
\left( \matrix{ \psi_0(k,r_2) \cr \psi'_0(k,r_2) \cr} \right)
   =\left( \matrix{ \psi_0(k,R) \cr \psi'_0(k,R) \cr} \right)
\ee
allows the determination of $A(r_2)$ (for $K_2^2 \leq 0$ the extension of
 the matricial equation is straightforward).
Indeed, the solution of this matricial equation yields
\be
A(r_2)=\frac{A(R)  -{\mathcal{T}} }{A(R) 
K_2^2 \ {\mathcal{T}} + 1} \ ,
\ee
with
\bea
{\mathcal{T}} & = & \frac{\tan(K_2 (R-r_2))}{K_2} 
\qquad\quad K_2=\sqrt{K_2^2} \nonumber\\
 &= & \frac{\tanh(K_2^* (R-r_2))}{K_2^*} \qquad\quad K_2^*=\sqrt{-K_2^2} \ .
\eea
To determine $K_1^2$ on the 
interval $[r_1,r_2[$ 
 use is made of  Eq. (\ref{solk1}) by setting $R=r_2$, namely
\be
K_1^2=\frac{A(r_2) B(r_2) + A(r_2) r_2 - B(r_2) r_2 - r_2^2}{A(r_2) 
B(r_2)  r_2^2} \ , 
\label{solk1p}
\ee
where $B(r_2) = r_2/2$.

Let us define
\be
F_0 = 54  A(R)  C(R) + 27  A(R) R - 54  C(R) R - 27 R^2 
\ee
\bea
F_1 & = &
- 27  A(R)  C(R) R^2 
 - 9  A(R)  R^3  +  9  C(R)  R^3 + 36  A(R)  C(R)  r_2^2 \nonumber\\ 
& & + 18  A(R)  R r_2^2 
  - 36  C(R)  R r_2^2 -  18  R^2 r_2^2 
 + 18  C(R)  r_2^3 + 9  R r_2^3 
\eea
\bea
F_2 & = &
 - 18  A(R)  C(R)  R^2 r_2^2 - 6  A(R)  R^3 r_2^2 
  + 6  C(R) R^3 r_2^2  \nonumber\\
& & + 18  A(R)  C(R)  R r_2^3 + 9  A(R)  R^2 r_2^3 
- 9  C(R)  R^2 r_2^3 -  3  R^3 r_2^3 
\eea
\be
 F_3 = 
- 3  A(R)  C(R)  R^3 r_2^3 
\ee
\bea
G_1 & = &
 - 18  A(R)  C(R) r_2^2  - 9  A(R) R r_2^2  
 + 18  C(R) R r_2^2  + 9 R^2 r_2^2  \nonumber\\
 & &  - 18  C(R) r_2^3  - 9 R r_2^3 
\eea
\bea
G_2 & = &
 9 A(R)C(R)  R^2 r_2^2 
 + 3  A(R)  R^3 r_2^2  - 3  C(R)  R^3 r_2^2 
  - 18  A(R)  C(R)  R r_2^3 \nonumber\\ 
 & & - 9  A(R)  R^2 r_2^3  
 + 9  C(R)  R^2 r_2^3 
 + 3  R^3 r_2^3  
 + 6  A(R)  C(R)  r_2^4\nonumber\\  
 & & + 3  A(R)  R r_2^4   
 - 6  C(R)  R r_2^4  - 3  R^2 r_2^4 
\eea
\be
 G_3 = + 3  A(R)  C(R)  R^3 r_2^3 
 - 3  A(R)  C(R)  R^2 r_2^4 
 -  A(R) R^3 r_2^4  +  C(R)  R^3 r_2^4 \ . 
\ee

We get the following equation
\be
F_0 + F_1 K_2^2 + F_2 K_2^4 + F_3 K_2^6 + G_1 K_1^2 + G_2 K_1^2K_2^2 +
G_3 K_2^4 K_1^2 = 0 \ . 
\ee
It gives $K_1^2$ in terms of $A(R),B(R),C(R)$ and $K_2^2, r_2$. 
Then $r_1$ is determined ($0 < r_1 < r_2$)  through Eq. (\ref{eqr1}). 
The value  of $K_0$ is such that 
$K_0 r_1=x_{1,n_1}$, whatever $n_1$, forcing  $r_1$ to be a zero of  $\psi_1$.

For scattering data  corresponding to a potential in the considered class,
 namely constant on intervals determined from zeros of the regular solution,
 the phase-equivalent potentials all correspond to a permutation of the zeros
of the regular solution with the constraint that 
$ (\forall \ell=1,\ldots,N-1) \quad r_{\ell,n_{\ell}} < r_{\ell+1,n_{\ell+1}}$
  and $r_{N,n_{N}} < R$.

The transparent potentials are zero on $[r_1,R[$. The values 
$r_1$ and $r_2$ are respectively 
zeros of the spherical   Bessel functions  $j_1$ and $j_2$.
 The potentials  are non-zero only on $[0,r_1[$.
 The value  of $K_0$, such that $K_0 r_1=x_1$ differs from $k=\sqrt{E}$ 
 if $r_1 \neq x_1/k$. This  is what leads to the discrete ambiguities.

{\bf Numerical Application}

We construct a reference potential, which is constant on each interval  
$[0,r_1[,[r_1,r_2[,[r_2,R[$. Here, we choose
 $v_0=-3, K_0=2, v_1=-8, K_1=3, v_2=-4, K_2=\sqrt{5}$, together with 
$r_1=2.24670598$ and $r_2=2.6958027$, which reduce  
 $\psi_1$ and $\psi_2$ to zero, respectively.
The phase-shifts $\delta_j,j=0,1,2$ of this potential are
calculated at $E=1$ by means of  $A(R),B(R),C(R)$.

Phase equivalent potentials are found by fitting piecewise constant
potentials to these  phase-shifts
 or more explicitly $A(R),B(R),C(R)$. 
First, the values of $K_2^2$ and $v_2$ are equal to the starting values  
$v_2=-4, K_2=\sqrt{5}$ according to Eq. (\ref{eqk2}) or Eq. (\ref{eqk2p}). 
Then, 
 $r_2$ is determined from  Eqs. (\ref{T22p0},\ref{T22p}) with the 
 constraint that $r_2 < R$. The value of $K_1$ is obtained from 
Eq. (\ref{solk1p}), and $r_1$ is determined ($0 < r_1 < r_2$)  
through Eq. (\ref{eqr1}). 
The value  of $K_0$ is such that 
$K_0 r_1=x_{1,n_1}$,  $r_1$ being a zero of  $\psi_1$. 

Once the phase equivalent potential is determined, the corresponding
Schr\"odinger equation is solved for each partial wave. It yields the
respective positions of the 
zeros of the regular solution. The resolution of Eq. (\ref{cal})
 yields a phase-shift  from  which one subtracts the phase-shift of 
 the reference potential. 

A sample of solutions for $R= 5.5$ and $R=10.0$ are displayed in Tables
3 and 4, respectively. 
The results underline a
classification according to three indices :

- $n$, the difference between the phase calculated directly from Eq. (\ref{cal})
 and the exact value divided by $\pi$.
 It is found to be the same for the three waves $\ell=0,1,2$.

- $n_1$, numbering the zero of regular solution of the $\ell= 1 $ wave,

- $n_2$, numbering the zero of the regular solution of the $\ell= 2$
wave.

The reference potential is recovered for the $\{n,n_1,n_2\}=\{0,1,1\}$ case.
Moreover, in each subclass of fixed $n$, the solution can be ordered
according to $\{n_1,n_2\}$ in such a way that $n_2 \geq n_1$.

\vspace*{.2cm}
\begin{table}
\begin{center}
\begin{tabular}{|r|r|r|r|r|r|r|}
\hline
$ n$ & $n_1$ & $n_2$ & $K_0$ & $r_1$ & $ K_1$ & $r_2$  \\
\hline
 -1&  1&  1&   1.1802&   3.8073&   3.9070&   4.1795 \\
\hline
\hline
  0&  1&  1&   2.0000&   2.2467&   3.0000&   2.6958 \\
\hline
\hline
  0&  1&  2&   1.4985&   2.9986&   3.9070&   4.1795 \\
  0&  2&  2&   2.0291&   3.8073&   3.9070&   4.1795 \\
\hline
  1&  1&  1&   8.9280&   0.5033&   2.6010&   0.8710  \\
  1&  1&  2&   3.8906&   1.1549&   3.0000&   2.6958 \\
  1&  1&  3&   2.0551&   2.1865&   3.9070&   4.1795 \\
  1&  2&  2&   3.4385&   2.2467&   3.0000&   2.6958 \\
  1&  2&  3&   2.5763&   2.9986&   3.9070&   4.1795 \\
  1&  3&  3&   2.8640&   3.8073&   3.9070&   4.1795 \\
\hline
  2&  1&  4&   3.2924&   1.3648&   3.9070&   4.1795 \\
  2&  2&  2&  15.3494&   0.5033&   2.6010&   0.8710 \\
  2&  2&  3&   6.6888&   1.1549&   3.0000&   2.6958 \\
  2&  2&  4&   3.5332&   2.1865&   3.9070&   4.1795 \\
  2&  3&  3&   4.8534&   2.2467&   3.0000&   2.6958 \\
  2&  3&  4&   3.6364&   2.9986&   3.9070&   4.1795 \\
 2&  4&  4&   3.6945&   3.8073&   3.9070&   4.1795 \\
\hline 
 3&  1&  5&   9.2930&   0.4835&   3.9070&   4.1795 \\
 3&  2&  5&   5.6605&   1.3648&   3.9070&   4.1795 \\
  3&  3&  3&  21.6656&   0.5033&   2.6010&   0.8710 \\
  3&  3&  4&   9.4412&   1.1549&   3.0000&   2.6958 \\
  3&  3&  5&   4.9870&   2.1865&   3.9070&   4.1795 \\
  3&  4&  4&   6.2608&   2.2467&   3.0000&   2.6958 \\
  3&  4&  5&   4.6909&   2.9986&   3.9070&   4.1795 \\
 3&  5&  5&   4.5231&   3.8073&   3.9070&   4.1795 \\
\hline
  4&  2&  6&  15.9769&   0.4835&   3.9070&   4.1795 \\
  4&  3&  6&   7.9897&   1.3648&   3.9070&   4.1795 \\
   4&  4&  4&  27.9484&   0.5033&   2.6010&   0.8710 \\ 
  4&  4&  5&  12.1791&   1.1549&   3.0000&   2.6958 \\
 4&  4&  6&   6.4332&   2.1865&   3.9070&   4.1795 \\
  4&  5&  5&   7.6649&   2.2467&   3.0000&   2.6958 \\
 4&  5&  6&   5.7429&   2.9986&   3.9070&   4.1795 \\
  4&  6&  6&   5.3506&   3.8073&   3.9070&   4.1795 \\
\hline
\end{tabular}
\end{center}
\caption{  Phase-equivalent potential to the reference potential of line  2. 
The range parameter is  $R=5.5$.
$K_j^2=k^2-v_j, k=1$. The classification is made according to indices $n,n_1,n_2$ 
(see text)}
\end{table}

\vspace*{.2cm}
\begin{table}
\begin{center}
\begin{tabular}{|r|r|r|r|r|r|r|}
\hline
$ n$ & $n_1$ & $n_2$ & $K_0$ & $r_1$ & $ K_1$ & $r_2$  \\
\hline
 -2&  1&  1&   0.8310&   5.4072&   7.1067&   5.6213 \\
\hline
 -1&  1&  1&   1.1802&   3.8073&   3.9070&   4.1795 \\
 -1&  1&  2&   0.9051&   4.9648&   7.1067&   5.6213 \\
 -1&  2&  2&   1.4287&   5.4072&   7.1067&   5.6213 \\
\hline
\hline
  0&  1&  1&   2.0000&   2.2467&   3.0000&   2.6958 \\
\hline
\hline 
 0&  1&  2&   1.4985&   2.9986&   3.9070&   4.1795 \\
  0&  1&  3&   0.9936&   4.5223&   7.1067&   5.6213 \\
  0&  2&  2&   2.0291&   3.8073&   3.9070&   4.1795 \\
  0&  2&  3&   1.5560&   4.9648&   7.1067&   5.6213 \\
  0&  3&  3&   2.0166&   5.4072&   7.1067&   5.6213 \\
\hline
   1&  1&  1&   8.9279&   0.5033&   2.6010&   0.8710 \\
  1&  1&  2&   3.8905&   1.1550&   3.0000&   2.6958 \\
  1&  1&  3&   2.0551&   2.1865&   3.9070&   4.1795 \\
 1&  1&  4&   1.1014&   4.0798&   7.1067&   5.6213 \\
   1&  2&  2&   3.4385&   2.2467&   3.0000&   2.6958 \\
 1&  2&  3&   2.5763&   2.9986&   3.9070&   4.1795 \\
  1&  2&  4&   1.7082&   4.5223&   7.1067&   5.6213 \\
  1&  3&  3&   2.8640&   3.8073&   3.9070&   4.1795 \\
  1&  3&  4&   2.1963&   4.9648&   7.1067&   5.6213 \\
 1&  4&  4&   2.6014&   5.4072&   7.1067&   5.6213 \\
\hline
  2&  1&  4&   3.2924&   1.3648&   3.9070&   4.1795 \\
  2&  1&  5&   1.2354&   3.6371&   7.1067&   5.6213 \\
  2&  2&  2&  15.3492&   0.5033&   2.6010&   0.8710 \\
  2&  2&  3&   6.6888&   1.1550&   3.0000&   2.6958 \\
  2&  2&  4&   3.5332&   2.1865&   3.9070&   4.1795 \\
  2&  2&  5&   1.8935&   4.0798&   7.1067&   5.6213 \\
  2&  3&  3&   4.8534&   2.2467&   3.0000&   2.6958 \\
  2&  3&  4&   3.6364&   2.9986&   3.9070&   4.1795 \\
  2&  3&  5&   2.4112&   4.5223&   7.1067&   5.6213 \\
  2&  4&  4&   3.6945&   3.8073&   3.9070&   4.1795 \\
  2&  4&  5&   2.8332&   4.9648&   7.1067&   5.6213 \\
  2&  5&  5&   3.1848&   5.4072&   7.1067&   5.6213 \\
\hline
\end{tabular}
\end{center}
\caption{ Same as Table 3 for a range parameter $R=10.0$. The
reference potential appears on line 5.}  
\end{table}

\newpage

\subsection{4 phase-shifts $\ell=0,1,2,3$}

We begin the procedure by determining the value of $K_3$ as  a
function of $r_3$. On the interval $[r_3,R[$, 
the regular solution $\psi_3$, which vanishes at $ r= r_3$, 
  is a  linear combination of 
$u_3(K_3 r)$ and $w_3(K_3 r)$ : 
\be
\psi_3(k,r)=u_3(K_3 r) w_3(K_3 r_3)-u_3(K_3 r_3) w_3(K_3 r) \ . 
\ee
According to the definition (\ref{ratioE}), 
the ratio $(\psi_3/\psi'_3)(k,R) = D(R)$ at  $r=R$. It yields
\bea
\frac{\tan(K_3 (R-r_3))}{K_3}& =& \frac{\mathcal{N}_3}{\mathcal{D}_3} 
\qquad\quad  \ K_3=\sqrt{ K_3^2}  \nonumber\\
\frac{\tanh(K_3^* (R-r_3))}{K_3^*}& =& \frac{\mathcal{N}_3}{\mathcal{D}_3} 
\qquad\quad K_3^*=\sqrt{-K_3^2}  \ ,
\label{T33}
\eea
where $\mathcal{N}_3, \mathcal{D}_3$ are given in Appendix {\bf A}.
We first  check if the potential is simply soluble. Looking at Eqs.(\ref{solk1},\ref{eqk2}),
we expect 
\be
K_3^2=\frac{9    C(R) D(R) + 3  C(R) R - 3  D(R) R - R^2}{   C(R) D(R) R^2}
\label{eqsK3}
\ee
or, equivalently, provided that $R$ is not a zero of  $\psi_2$ or $\psi_3$
\bea
K_3^2&=&k^2+\frac{\mathcal{N}}{\mathcal{D}} \nonumber\\
\mathcal{N}&=&k^7 R^5 \sin(\delta_2 - \delta_3) \nonumber\\ 
\mathcal{D} & = & ((3 - k^2 R^2) \sin(k R + \delta_2)-
                     3 k R \cos(k R + \delta_2)) \nonumber\\
              & \times & ( 3   (5 - 2  k^2  R^2) \sin(k R + \delta_3)    
	      +k  R  ( k^2  R^2-15)   \cos(k R + \delta_3)) \ .
\eea  
The equation (\ref{eqsK3}) is verified if and only if  
$(\psi_{2}/\psi'_{2})(r=r_3)=C(r_3)=r_3/3$. This can be proved
in the following way. On the interval  $[r_3,R[$, we
express $\psi_2$ as
\be
\psi_2(k,r)=  \alpha(K_3) u_2(K_3 r) + \beta(K_3)  w_2(K_3 r) \ .
\ee
The ratio 
$\alpha(K_3)/\beta(K_3)$ is determined in terms of $C(r_3)$ and $C(R)$, 
which yields  
\bea
\frac{\tan(K_3 (R-r_3))}{K_3}& =& \frac{\mathcal{N'}_3}{\mathcal{D'}_3} 
\qquad\quad  \ K_3=\sqrt{ K_3^2}  \nonumber\\
\frac{\tanh(K_3^* (R-r_3))}{K_3^*}& =& \frac{\mathcal{N'}_3}{\mathcal{D'}_3} 
\qquad\quad K_3^*=\sqrt{-K_3^2} \ ,
\label{T22n}
\eea
where ${\mathcal{N'}_3}$ and ${\mathcal{D'}_3}$ are 
given in Appendix {\bf A}.

By equating $\mathcal{N}_3/\mathcal{D}_3 = \mathcal{N'}_3/\mathcal{D'}_3$, and
taking $K_3^2$ from (\ref{eqsK3}), we find indeed $C(r_3) = r_3/3$.

It remains to actually calculate $C(r_3)$. To do so, we first note that
$D(r_1)=5 r_1/(x_{1,n_1}^2-15)$.  Here, use is made of the fact that 
$K_0r_1 = x_{1,n_1}$ is a zero of $\psi_1$ and thus
$\sin(K_0r_1) = K_0r_1 \cos(K_0 r_1)$.

On the interval $[r_1,r_2[$, $\psi_3$ is written as 
\be
\psi_3(k,r)=  \alpha(K_1) u_3(K_1 r) + \beta(K_1)  w_3(K_1 r) \ .
\ee
The ratio $\alpha(K_1)/\beta(K_1)$ is given 
as a function of $D(r_1)$, thus as a function of $r_1, x_{1,n_1}$.
This ratio being known, $D(r_2)$ can be calculated.

Taking into account Eq.(\ref{eqr1}), $D(r_2)$ is rational fraction 
in terms  of $x_{1,n_1},K_1^2,$ $r_1,r_2$.
 
Finally, on $[r_2,r_3[$, $\psi_3$ is given by 
\be
\psi_3(k,r)=  \alpha(K_2) u_3(K_2 r) + \beta(K_2)  w_3(K_2 r) \ .
\label{eq56}
\ee
Similarly to the preceding step, the  ratio 
$\alpha(K_2)/\beta(K_2)$ is a function of $D(r_2)$. 
Inserting its value in (\ref{eq56}), and by using Eqs. (\ref{T22mp}) and (\ref{T22mpp}) at $R=r_3$, 
we obtain a compact polynomial expression  involving   
$x_{1,n_1}, K_1^2,K_2^2,r_1,r_2,r_3,C(r_3)$. By remembering that
$D(r_3) = 0$ and setting 
$C(r_3)=r_3/3$, the condition for a simple soluble model reads
\be
(9 + 3 K_1^2  r_2^2 + K_1^4  r_2^4) 
(225 + 45 K_2^2 r_2^2 + 6 K_2^4 r_2^4 + K_2^6 r_2^6) 
 (K_1^2  r_1^2 - x_{1,n_1}^2)=0 \ .
\ee

Setting $K_1^2r_2^2 = 3 x$, $K_2^2r_2^2 = y$ and $K_1 r_1 = z$, the
polynomial in $x$ has no real root, the polynomial in $y$ has a single
real root, namely $K_2^2 r_2^2=  -5.39254.$ 
 The polynomial in $z$ has 
root  $K_1 r_1=x_{1,n_1}$ ($K_1^2 \geq 0$)
implying $K_0 = K_1$. Consequently, $C(r_3) = r_3/3$ occurs only
in exceptional situations, and the model is not simply soluble.

Since the model is not simply soluble we let $r_3$  run  
from $0$ to $R$. Use is made of
Eq. (\ref{T33}) 
  and  the first equation of Appendix {\bf A} to determine $K_3$. 
The value $K_3 (R-r_3)$  is determined in intervals
 $]m_3 \pi, (m_3+1) \pi [$, where $m_3$ ($m_3 \geq 0$) is fixed at each run. 
 We have ensured that a single solution exists in
 each interval, with $K_3 $ 
being a continuous function of $r_3$ and $\psi_3(r_3) = 0$. Then
 $A(r_3),B(r_3),C(r_3)$  (see Eq. (\ref{ratioE})) for $\ell=0,1,2$, 
 respectively, are determined 
 from $A(R),B(R),C(R)$ by using the fact that on $[r_3,R[$ we have
\be
\psi_{\ell}(k,r)=\alpha_{\ell} u_{\ell}(K_3 r) + \beta_{\ell} w_{\ell}(K_3 r) \ .
\ee
By comparison with the previous case of 3 phase shifts, for which
$K_1,K_2$ are independent on $r_1,r_2$ respectively, many more solutions have to be considered.

We first look at transparent potentials,
for which the exact phase is $n \pi$ for all waves $\ell=0,1,2,3$. \footnote{Note that here we 
  may encounter transparent potentials for which the 
 fourth  exact phases are  $n(\ell) \pi$, 
with $n(\ell)$ not the same for all the waves.}
 The results are displayed in Tables  5 and  6 for 
 $R=12$ and $R=15$, respectively.
 
We denote by $n$ the class of phase-equivalent potential, for which 
 the exact phase  
 is $n \pi$ for all waves $\ell=0,1,2,3$. The present study emphasizes 
 these classes.  
 Each class is characterized by the set of numbers
 $\{n_1,n_2,n_3\}$ where $n_{\ell}$ is 
 $n_{\ell}$-th zero of the regular solution of the Schr\"odinger equation 
 for the $\ell$-wave.
The configurations can be ordered according to 
$\{n_1,n_2,n_3 \}$ with $ n_1 \leq n_2 \leq n_3 \leq n_{max}$. The number
of configurations and $n_{max}$ for fixed $n$ increase with $R$.

Obviously, except for the lowest exact phase, 
different sets  of  indices $n_1 \leq n_2 \leq n_3$ exist,
with different  $K_j,v_j$ giving the same phase $n \pi$. 
Consequently, the exact phase is not a sufficient criterion 
 to remove the phase-ambiguities, as suggested in ref.\cite{DSB}.
The size of the set  associated to the index $n$ increases with the 
value of $n$.

Obviously, for each $n$, the set of minimal size 
corresponds to 
$\{n_1,n_2,n_3\}=\{1,1,1\} ; (n_{max}=1)$. For the lowest $n$, this
set is the only solution. Note that in Table 5 and 6 for $n = 0$, 
all the solutions reduce to the same zero potential.

\vspace*{.2cm}
\begin{table}
\begin{center}
\begin{tabular}{|r|r|r|r|r|r|r|r|r|r|r|r|}
\hline
$ n$ & $n_1$ & $n_2$ & $n_3$ & $K_0$ & $r_1$ & $ K_1$ & $r_2$&  $K_2$ & $r_3$ & $K_3$ & $R$  \\
\hline
 -1& 1 & 1 & 1&   0.808&   5.561&   0.736&   7.259&   0.607&   9.703&   0.714&  12.000 \\
 \hline
 0& 1 & 1 & 1 &    1.000&   4.493&   1.000&   5.763&   1.000&   6.988&   1.000&  12.000 \\
   0& 1 & 1 & 2 &   1.000&   4.493&   1.000&   5.763&   1.000&  10.417&   1.000&  12.000 \\
  0& 1 & 2 &2 &   1.000&   4.493&   1.000&   9.095&   1.000&  10.417&   1.000&  12.000 \\
    0&  2 & 2 & 2 &  1.000&   7.725&   1.000&   9.095&   1.000&  10.417&   1.000&  12.000 \\
\hline
  1&  1 & 1 & 1 &  1.141&   3.936&   1.249&   4.968&   1.483&   5.716&   1.310&  12.000 \\
  1& 1 & 1 & 2 &    1.171&   3.838&   1.251&   4.864&   1.398&   8.081&   1.240&  12.000 \\
  1& 1 & 1 & 3 &   1.174&   3.827&   1.237&   4.863&   1.340&  10.681&   1.191&  12.000 \\
 1& 1 & 2 & 2 &   1.155&   3.892&   1.264&   7.530&   1.538&   8.316&   1.313&  12.000 \\
  1& 1 & 2 & 3 &    1.161&   3.870&   1.240&   7.579&   1.423&  10.728&   1.233&  12.000 \\
   1&  1 & 3 & 3 &  1.160&   3.872&   1.272&  10.021&   1.599&  10.812&   1.318&  12.000 \\
   1& 2 & 2 & 2 &   1.119&   6.901&   1.355&   7.933&   1.765&   8.616&   1.421&  12.000 \\
    1& 2 & 2 & 3 &   1.135&   6.805&   1.292&   7.881&   1.537&  10.784&   1.288&  12.000 \\
  1& 2 & 3 & 3 &   1.126&   6.863&   1.367&  10.231&   1.853&  10.907&   1.430&  12.000 \\
   1& 3 & 3 & 3 &  1.079&  10.102&   1.878&  10.889&   3.562&  11.232&   2.029&  12.000 \\
\hline
\end{tabular}
\end{center}
\caption{ Transparent potentials such that the first four  phase-shift are
all equal to  $n \pi$. $K_j^2=k^2-v_j; \ k=1$. The range of the potential 
$R=12$.
The index $n_{\ell}$ is such that $r_{\ell}$ 
 is the $n_{\ell}$-th zero of the regular solution in the $\ell$-wave.}
\end{table}

\vspace*{.2cm}
\begin{table}
\begin{center}
\begin{tabular}{|r|r|r|r|r|r|r|r|r|r|r|r|}
\hline
$ n$ & $n_1$ & $n_2$ & $n_3$ & $K_0$ & $r_1$ & $ K_1$ & $r_2$&  $K_2$ & $r_3$ & $K_3$ & $R$   \\
\hline
   -2& 1 & 1 & 1 &   0.641&   7.008&   0.538&   9.297&   0.456&  13.285&   0.705&  15.000   \\
 \hline
   -1& 1 & 1 & 1 &   0.833&   5.396&   0.756&   7.048&   0.683&   9.210&   0.803&  15.000   \\
   -1& 1 & 1 & 2 &   0.835&   5.384&   0.781&   6.992&   0.740&  13.542&   0.869&  15.000   \\
    -1&1 & 2 & 2 &    0.833&   5.391&   0.772&  11.360&   0.699&  13.512&   0.847&  15.000   \\
   -1& 2 & 2 & 2 &    0.847&   9.117&   0.713&  10.998&   0.632&  13.458&   0.810&  15.000   \\
\hline 
    0& 1 & 1 & 1 &    1.000&   4.493&   1.000&   5.763&   1.000&   6.988&   1.000&  15.000   \\
    0& 1 & 1 & 2 &    1.000&   4.493&   1.000&   5.763&   1.000&  10.417&   1.000&  15.000   \\
    0& 1 & 1 & 3 &   1.000&   4.493&   1.000&   5.763&   1.000&  13.698&   1.000&  15.000   \\
    0& 1 & 2 & 2 &    1.000&   4.493&   1.000&   9.095&   1.000&  10.417&   1.000&  15.000   \\
     0& 1 & 2 & 3 &   1.000&   4.493&   1.000&   9.095&   1.000&  13.698&   1.000&  15.000   \\
     0& 1 & 3 & 3 &    1.000&   4.493&   1.000&  12.323&   1.000&  13.698&   1.000&  15.000   \\
     0& 2 & 2 & 2 &   1.000&   7.725&   1.000&   9.095&   1.000&  10.417&   1.000&  15.000   \\
   0& 2 & 2 & 3 &    1.000 &   7.725 &   1.000&   9.095&   1.000&  13.698&   1.000&  15.000   \\
   0&  2 & 3 & 3 &  1.000&   7.725&   1.000&  12.323&   1.000&  13.698&   1.000&  15.000   \\
    0&  3 & 3 & 3 &  1.000&  10.904&   1.000&  12.323&   1.000&  13.698&   1.000&  15.000   \\
\hline 
 1& 1 & 1 & 1 &   1.120&   4.013&   1.234&   5.058&   1.395&   5.831&   1.229&  15.000   \\
 1 & 1 & 1 & 2 &   1.155&   3.892&   1.240&   4.928&   1.328&   8.307&   1.176&  15.000   \\
 1& 1 & 1 & 3 &    1.160&   3.873&   1.227&   4.917&   1.284&  10.987&   1.139&  15.000   \\
  1& 1 & 1 & 4 &   1.154&   3.892&   1.207&   4.952&   1.247&  13.807&   1.111&  15.000   \\
  1& 1 & 2 & 2 &    1.142&   3.935&   1.244&   7.630&   1.393&   8.485&   1.206&  15.000   \\
  1& 1 & 2 & 3 &   1.151&   3.902&   1.227&   7.651&   1.320&  11.044&   1.155&  15.000   \\
   1& 1 & 2 & 4 &    1.148&   3.915&   1.205&   7.731&   1.269&  13.815&   1.121&  15.000   \\
   1& 1 & 3 & 3 &    1.149&   3.912&   1.239&  10.227&   1.382&  11.136&   1.183&  15.000   \\
   1&  1 & 3 & 4 &  1.145&   3.924&   1.211&  10.386&   1.306&  13.829&   1.136&  15.000     \\
   1& 1 & 4 & 4 &    1.148&   3.915&   1.226&  12.896&   1.366&  13.850&   1.161&  15.000   \\
  1& 2 & 2 & 2 &  1.109&   6.967&   1.317&   8.027&   1.524&   8.803&   1.264&  15.000   \\
  1& 2 & 2 & 3 &    1.127&   6.857&   1.272&   7.948&   1.392&  11.150&   1.187&  15.000   \\
   1& 2 & 2 & 4 &    1.128&   6.846&   1.235&   7.966&   1.313&  13.832&   1.139&  15.000   \\
   1& 2 & 3 & 3 &    1.119&   6.902&   1.299&  10.446&   1.491&  11.281&   1.229&  15.000   \\
  1& 2 & 3 & 4 &    1.123&   6.879&   1.247&  10.567&   1.366&  13.850&   1.161&  15.000   \\
   1&  2 & 4 & 4 &  1.122&   6.886&   1.275&  12.991&   1.456&  13.880&   1.197&  15.000   \\
    1&  3 & 3 & 3 &  1.090&  10.003&   1.476&  10.988&   1.840&  11.651&   1.365&  15.000   \\
    1& 3 & 3 & 4 &    1.103&   9.889&   1.337&  10.967&   1.520&  13.899&   1.223&  15.000   \\
    1& 3 & 4 & 4 &   1.097&   9.943&   1.412&  13.218&   1.720&  13.955&   1.299&  15.000   \\
     1& 4 & 4 & 4 &    1.055&  13.338&   2.302&  13.994&   4.102&  14.284&   1.993&  15.000   \\
\hline
\end{tabular}
\end{center}
\caption{ Same as Table 5 but for $R=15$}
\end{table}

As a more realistic example, the reference potential 
is chosen as the starting potential with parameters $E=k^2=1$, $R=10$,
\bea
\sqrt{k^2-v_0}=K_0 & = & 2 \qquad\quad   0 \leq r < r_1= 2.24668994 \nonumber\\
\sqrt{k^2-v_1}=K_1 & = & 3 \qquad\quad   r_1 \leq r < r_2= 2.699578766 \nonumber\\
\sqrt{k^2-v_2}=K_2 & = & \sqrt{5} \qquad\quad   r_2 \leq r < r_3= 7.38655698 \nonumber\\
\sqrt{k^2-v_3}=K_3 & = & \sqrt{2} \qquad\quad   r_3 \leq r <  10  \ .
\eea
The functions $A(R),B(R),C(R),D(R)$ are calculated, and 
the procedure developed in 
 subsection {\bf 3.4}   is used to determine phase-equivalent potentials.

The classification of the solutions is performed very much in the same
way as for transparent potentials. Here $n$ is the difference between
the exact phases of the reference and the phase-equivalent potentials
divided by $\pi$.
 A priori, $n$ depends on $\ell$. As an example, the following
 phase-equivalent potential $v_j=k^2-K_j^2$
\bea
 K_0,r_1 & = &  3.12729631, 1.43683507 \nonumber\\
 K_1,r_2 & = &  1.39601562, 4.82494432  \nonumber\\
 K_2,r_3 & = & 2.54346784,  4.97266696 \nonumber\\
 K_3,R & = &  0.820438746,  10
\eea
is such that the first 3 values of $n(\ell=0)$ = $n(\ell=1)$ =
$n(\ell=2)$ = -2, while $n(\ell =3)$ = -3. In this case, the 3 indices
$n_{\ell}$ are not ordered according to
$n_1 \leq n_2 \leq n_3$. Actually, 
$n_1=1, n_2=2,n_3=1$.  

Let us concentrate on potentials for which $n$ is independent  of
$\ell$. Results are displayed in Table 7. The same
conclusions apply to this ensemble as to the case of transparent
potentials. We note that one of the $n=0$ solutions reproduces the
reference potential.

\vspace*{.2cm}
\begin{table}
\begin{center}
\begin{tabular}{|r|r|r|r|r|r|r|r|r|r|r|r|}
\hline
$ n$ & $n_1$ & $n_2$ & $n_3$ & $K_0$ & $r_1$ & $ K_1$ & $r_2$&  $K_2$ & $r_3$ & $K_3$ & $R$   \\
\hline
  -3& 1 & 1 & 1 &    1.007&   4.464&   1.825&   5.219&   2.072&   5.479&   0.877&  10.000 \\
\hline
 -2& 1 & 1 & 1 &   1.267&   3.546&   2.222&   4.164&   2.308&   4.391&   1.294& 10.000  \\
  -2& 1 & 1 & 2 &   1.364&   3.295&   2.200&   3.913&   1.996&   5.856&   0.934&  10.000 \\
 -2& 1 & 2 & 2 &    1.339&   3.355&   2.174&   5.469&   2.004&   5.771&   0.920&  10.000 \\
   -2& 2 & 2 & 2 &    1.497&   5.161&   2.621&   5.712&   1.990&   6.042&   0.968&  10.000 \\
\hline
      -1&1 & 1 & 1 &    1.464&   3.069&   2.697&   3.594&   2.541&   3.782&   2.847&10.000 \\
    -1& 1 & 1 & 2 &    1.635&   2.749&   2.657&   3.261&   2.278&   4.959&   1.413& 10.000 \\
    -1& 1 & 1 & 3 &   1.771&   2.538&   2.640&   3.048&   2.044&   6.597&   1.099&  10.000 \\
     -1& 1 & 2 & 2 &   1.626&   2.763&   2.627&   4.513&   2.274&   4.769&   1.369&10.000 \\
   -1& 1 & 2 & 3 &    1.781&   2.523&   2.626&   4.273&   1.999&   6.235&   1.008&  10.000 \\
   -1&1 & 3 & 3 &    1.710&   2.627&   2.553&   5.680&   1.990&   6.006&   0.961&  10.000 \\
    -1 & 2 & 2 & 2&   1.756&   4.400&   3.146&   4.860&   2.290&   5.130&   1.457&  10.000 \\
   -1& 2 & 2 & 3 &    1.884&   4.100&   3.055&   4.568&   2.031&   6.510&   1.075&  10.000 \\
   -1& 2 & 3 & 3 &    1.800&   4.291&   2.955&   5.853&   1.996&   6.193&   0.999&  10.000 \\
  -1& 3 & 3 & 3 &    1.954&   5.581&   3.435&   6.011&   2.011&   6.359&   1.037&  10.000 \\
 \hline 
    0&1 & 1 & 1 &    1.617&   2.779&   3.034&   3.235&   2.752&   3.390&   2.073&  10.000 \\
    0& 1 & 1 & 2 &   1.782&   2.522&   3.076&   2.967&   2.544&   4.464&   1.869&  10.000 \\
   0& 1 & 1 & 3 &    1.905&   2.359&   3.071&   2.801&   2.389&   5.798&   1.670& 10.000 \\
\hline
\hline
  0&  1 & 1 & 4 &  2.000&   2.247&   3.000&   2.696&   2.236&   7.387&   1.414&  10.000 \\
\hline
\hline
     0&1 & 2 & 2 &    1.813&   2.478&   3.056&   3.982&   2.529&   4.191&   1.789&  10.000   \\
    0& 1 & 2 & 3 &    2.009&   2.236&   3.094&   3.721&   2.320&   5.391&   1.531&10.000 \\
  0& 1 & 2 & 4 &    2.181&   2.061&   3.081&   3.552&   2.118&   6.960&   1.220&  10.000 \\
   0&1 & 3 & 3 &    2.004&   2.242&   3.046&   4.801&   2.285&   5.069&   1.441&10.000 \\
 0 & 1 & 3 & 4 &   2.211&   2.033&   3.067&   4.576&   2.032&   6.517&   1.077&  10.000 \\
  0& 1 & 4 & 4 &    2.105&   2.135&   2.953&   5.853&   1.996&   6.193&   0.998&  10.000 \\
  0& 2 & 2 & 2 &   1.921&   4.021&   3.674&   4.417&   2.561&   4.634&   1.923&  10.000 \\
 0& 2 & 2 & 3 &    2.066&   3.739&   3.615&   4.139&   2.382&   5.763&   1.656&  10.000 \\
0 & 2 & 3 &  3&   2.063&   3.744&   3.521&   5.057&   2.312&   5.331&   1.513&  10.000 \\
   0& 2 & 2 & 4 &    2.205&   3.503&   3.504&   3.912&   2.182&   7.208&   1.324&  10.000 \\
  0& 2 & 3 & 4 &    2.239&   3.450&   3.450&   4.784&   2.063&   6.700&   1.130&  10.000 \\
   0&  2 & 4 & 4 &  2.131&   3.625&   3.313&   5.975&   2.007&   6.322&   1.028&  10.000 \\
   0 & 3 & 3& 3 &    2.207&   4.941&   4.118&   5.301&   2.348&   5.576&   1.590&  10.000 \\
  0& 3 & 3 & 4 &    2.360&   4.621&   3.940&   4.995&   2.099&   6.879&   1.190&  10.000 \\
0& 3 & 4 & 4 &   2.242&   4.864&   3.763&   6.096&   2.022&   6.447&   1.059&  10.000 \\
  0& 4 & 4 & 4 &    2.405&   5.848&   4.267&   6.199&   2.037&   6.553&   1.087&  10.000 \\
\hline
\end{tabular}
\end{center}
\caption{ Same as Tables  5 and  6 but for a non zero starting potential
exhibited between the double lines.}
\end{table}

The configuration $n=1$ for non transparent potentials is reported in Appendix {\bf B}.

\section{Generalization by means of the JWKB approximation}

The method developed in the preceding sections is not suitable for an
extension to a larger number of phase shifts, as it would  require a
tremendous numerical effort. In order to work beyond 5 phase-shifts 
 and obtain  some general results, use can be made of the JWKB
approximation. We recall that in this case the phase-shift is given by:
\be
       \tilde \delta(\ell,k)=\lim_{r \to \infty} 
        \bigg(\int_{rt(\ell,k)}^r K(\ell,k,r') \  dr'
 - \int_{\tilde{rt}(\ell,k)}^r  
            \tilde{K}(\ell,k,r') dr' \bigg) 
\label{phase}
\ee
(remember that  $E=k^2$). 
For the type of potential considered in this work, we have
\be
      K(\ell,k,r)=\sqrt{k^2-v(r)-\frac{\ell (\ell+1)}{r^2} } \qquad\quad      
      \tilde{K}(\ell,k,r)=\sqrt{k^2-\frac{\ell (\ell+1)}{r^2}}
\ . \ee
In (\ref{phase}),  $rt(\ell,k)$ is ``the'' largest turning point 
relative to the function $K$ considered, while 
$\tilde{rt}(\ell,k) = \sqrt{\ell(\ell+1)}/k$.

The JWKB phase $\tilde{\delta}(\ell,k)$ has been calculated for the 
transparent potentials of Table  5, and compared to the exact phase.
In case of the null potential, the JWKB estimate is obviously exact.
For non zero potentials, the relative differences 
\be
D_{\ell} = \vert \frac{\delta(\ell,k)-\tilde \delta(\ell,k)}{\delta(\ell,k)} \vert 
\ee
are most of the time better than 1 \%,  but as large as  2 \% in
a few instances.

Furthermore, the JWKB method can be shown to predict the position of the 
zeros of the regular solution.
To this aim, the regular solution is written
\be
\psi_{\ell}(k,r) \propto \sin \left( \int_{rt(\ell,k)}^r \ 
\sqrt{k^2-v(r')-\frac{\ell (\ell+1)}{r'^2} } \ dr' \right) \ .
\ee

Clearly, for
 $\hat r_n(\ell,k)$ to be the nth zero of the regular solution, the
 following condition has to be satisfied :
\be
\int_{rt(\ell,k)}^{\hat r_n(\ell,k)}  \ 
\sqrt{k^2-v(r')-\frac{\ell (\ell+1)}{r'^2} } \ dr' = n \pi \ .
\ee
Actually, the quantity to be considered is 
\be
I_{\ell}=\frac{1}{\pi} \left(\int_{rt(\ell,k)}^{\hat r_n(\ell,k)}  \ 
\sqrt{k^2-v(r')-\frac{\ell (\ell+1)}{r'^2} } \ dr' \right)+c_{\ell} \ .
\ee
The levelling value $c_{\ell}$ is given by
\be
c_{\ell}  = 1- \frac{1}{\pi} 
\int_{\sqrt{\ell (\ell+1)}/k}^{r_1(\ell)/k}  \ 
\sqrt{k^2-\frac{\ell (\ell+1)}{r'^2} } \ dr'  \ ,
\label{cl}
\ee
where $r_1(\ell)$ is the first zero of the spherical Bessel $j_{\ell}(z)$. 
The quantity $c_{\ell}$ does not depend on the value of $k$. It is
designed in a way that here 
$I_{\ell}$ is an integer and yields the exact result 
1 for the first zero in the
absence of potential.

Calculations have been performed for the transparent potentials of
Table 5. The ratio between the exact values and the JWKB estimates
show an excellent agreement. The deviation with respect 
to unity is most of the time better than 1 \%, with
few exceptions of the order of a few \%. In this case, $I_{\ell}$ is
close to an 
integer and represents the position of the $n$th zero of the regular
solution for the $\ell$ wave.

Comforted by the good quantitative agreement of the above calculations, 
the JWKB approximation can be used to ask questions on a general basis. 
Note that $N \geq 3$ is required, otherwise the model is
simple soluble and the  JWKB approximation is not necessary.
For instance, it is possible to predict the lowest value of  
 $n$ for a set of $N+1$ phase-shifts, and a fixed potential range $R$. 
 According to the results displayed in the previous Tables,
 it occurs for the configuration in which all the $n_N = 1$. Thus, it
 is sufficient to consider the largest value of $N$. 

 The data are the number of phase
shifts $N+1$ the energy $E=k^2$  and the potential range $R$.
Recalling the previous conventions,  ${\hat{r}_1(N,k)}$ denotes the first zero 
of the regular solution for the $N$-wave, supposed to be strictly lower than $R$.

We first consider transparent potentials. By definition, we have
\be
n = \frac{Q_1 + Q_2 - Q_3}{\pi} \ , \ee
with
\be
Q_1 =  \int_{rt(N,k)}^{\hat{r}_1(N,k)}  \ 
\sqrt{k^2-v(r')-\frac{N (N+1)}{r'^2} } \ dr' \ , \ee
\be
Q_2 =\int_{\hat{r}_1(N,k)}^R  \ \sqrt{k^2-v(r') -\frac{N (N+1)}{r'^2} } dr' \ ,
\ee
and
\be
 Q_3 = \int_{\sqrt{N (N+1)}/k}^{R}  \ \sqrt{k^2-\frac{N (N+1)}{r'^2} } dr' 
 \ .   \label{ninf}
\ee
The first contribution yields
\be
Q_1 = \pi- c_N \pi \ .
\ee
However, $c_N$ is defined by Eq.(\ref{cl}),  which yields
\be
Q_1 =  \int_{\sqrt{N (N+1)}/k}^{r_1(N)/k}  \ 
\sqrt{k^2-\frac{N (N+1)}{r'^2} } \ dr'  \ .
\label{Q1N}
\ee
The potential is constant on the interval 
$[\hat r_{1}(N,k),R[$ so that the two next integrals are analytical
and read
\bea  
Q_2 & = & 
\sqrt{K_N^2 R^2 -N(N+1)} - \sqrt{N(N+1)}
\arctan{ \sqrt{ \frac{K_N^2 R^2 - N(N+1)}{N(N+1)} }} \\\nonumber
& - & 
\sqrt{K_N^2\hat r_{1}^2(N,k)  -N(N+1)} - \sqrt{N(N+1)}
\arctan{ \sqrt{ \frac{K_N^2\hat r_{1}^2(N,k)  - N(N+1)}{N(N+1)} }} \\\nonumber
\eea
and
\be
Q_3 = \sqrt{k^2R^2 - N(N+1)} - \sqrt{N(N+1)} 
\arctan{ \sqrt{\frac{k^2 R^2 - N(N+1)}{N(N+1)} } }  \ .
\ee

It is clear that for the configurations having all $n_N = 1$, the value
of $n$ is ordered by the value of $\hat r_{1}(N,k)$. The lowest 
configuration corresponds
to $\hat r_{1}(N,k)$ being the closest to $R$. The absolute limit is
given, provided  $k R$  high enough and $ R \ne \hat r_{1}(N,k) $, by  
\be
n_{\rm min} >   \frac{Q_1 - Q_3}{\pi} \ , \ee
because $Q_2$ is strictly  positive.  In a compact form taking into account 
 the equation (\ref{Q1N}) we obtain
\be
n_{\rm min} >   -\frac{1}{\pi} 
\int_{r_1(N)/k}^{R}  \ \sqrt{k^2-\frac{N (N+1)}{r'^2} } dr' \ . 
\label{nmina}
\ee
Actually, $n_{\rm min}$ is integer and we have to take 
\be
n_{\rm min} \geq 1+\left[- \frac{1}{\pi} 
\int_{r_1(N)/k}^{R}  \ \sqrt{k^2-\frac{N (N+1)}{r'^2} } dr' \right] \ .
\label{nminb}
\ee
 Here, $[x]$ denotes the integer part of $x$, i.e.  the integer $m_x$ such that $m_x \leq x < m_x +1$. 
 In the present case we use the property  that any integer $m$ which satisfies 
$m > x$ satisfies $ m \geq [x]+1$.

If we take as an example the upper case of Tables  5 and 6, the numerical
estimate of Eq.(\ref{nmina}) yields   
$n_{\rm min}  > -1.47631 $ and  
$n_{\rm min}  > -2.39885 $ for $R=12$
 and $R=15$, respectively. 
 The integer values satisfy   $n_{\rm min} \geq -1$  and $n_{\rm min}
 \geq -2$, respectively,  
in agreement with the values of Tables 5 and 6 for $R=12$ and $R=15$.

On the other hand, by looking at the Tables 5,6 and 7, the maximum
value of $n_3$ depends on $n$. It obeys the following
 recurrence relation
\be
n_{3,{\rm max}}( n + 1) = n_{3,{\rm max}}(n) + 1 \ . \ee

For the transparent potentials,  $n_{3,{\rm max}}(0)$ is given by the number of 
zeros of the regular spherical function $j_3(kr)$  below $k R$  
 (we exclude the exceptional situation where a zero of $j_3$ is equal to $k R$).
 For instance let us consider $R=12,k=1$.
There are two zeros of $j_3$  below $12$, namely $6.988$ and $10.417$.
 They can be found in the   Table  5 concerning  the configuration 
 $n=0$ (zero potential). This implies that $n_{3,{\rm max}}(0)=2$ so that 
  $n_{3,{\rm max}}(-1)=n_{3,{\rm max}}(0)-1=-1$ (see Table 5).
  
A similar example is found from Table  6. Three zeros of 
$j_3$ are found below $k R=15$ : 
$6.988, 10.417$ and $13.698$. It yields  $n_{3,{\rm max}}(0)=3$, 
 which implies that  $n_{3,{\rm max}}(-2)=1$. We recover the lowest configuration 
 of the Table  6.

Yet a further example is provided us by $R=21, k=1$. Five zeros of $j_3$ 
exist below $21$ : 
  $6.988, 10.417,13.698, 16.924$ and $20.122$.
 It gives  $n_{3,{\rm max}}(0)=5$ and 
 $n_{3,{\rm max}}(-4)=1$. This configuration has been confirmed numerically.
It corresponds to the potential ($v_j=k^2-K_j^2=1-K_j^2$):
\bea
 K_0,r_1 &=  & 0.436410526, 10.2962892 \nonumber\\
 K_1, r_2 &= & 0.34393, 13.836594  \nonumber\\
 K_2, r_3 & = &  0.321277,  20.0371883 \nonumber\\
 K_3,R & = & 0.791895048, 21  \ .
\eea
The exact phases of this potential are $-4 \pi$ for $\ell=0-3$.

Taking the same case, $R=21, k=1$, but for  5 phases instead of 4, 
 4 zeros of $j_4$ are situated below $21$, namely 
$8.183,11.705, 15.040$ and $18.301$.
As expected, $n_{4,{\rm max}}(0)=4$ and  
$n_{4,{\rm max}}(-3)=1$. The lowest configuration is given by the following
potential ($v_j=k^2-K_j^2=1-K_j^2$):
\bea
K_0,r_1 & = &  0.536260007,  8.37916127 \nonumber\\
K_1, r_2 & = &   0.218548173,  11.03235 \nonumber\\
K_2,r_3 & = &   0.360733527,  13.5849 \nonumber\\
K_3,r_4 & = &   0.793613806,  14.7336 \nonumber\\
K_4,R & = &   0.266900301,  21. \nonumber\\
\eea
Its exact phases are  $-3 \pi$ for $\ell=0-4$

Beyond 5 phases, finding numerical solutions becomes too
tedious to be interesting.

 Taking into account the results for 4 and 5 phases 
($\ell=0,1,2,3$ and $\ell=0,1,2,3,4$), we propose the following 
 conjecture :
 
Consider models which are not simply soluble, i.e. number of phases $N + 1$
strictly  greater than 3, and  configurations such  that $n$
is the same for all partial waves, the minimal value of $n$ is given by 
 $n_{\rm min} = -n(j_N)+1$, where 
$n(j_N)$ denotes the number of zeros of the regular spherical Bessel function
 $j_N$ strictly smaller than   $k R$,   $k R$    not being
 a zero of $j_N$ and high enough to ensure $ n(j_N) \ne 0$. 

As stated in the preceding section, our attention has been
 focused on  cases such that  the phase difference is $n \pi$, independently of $\ell$.
 However, phase equivalent potentials with $n$ depending on $\ell$ can
be constructed. Two examples are quoted below for transparent potentials.

The first example corresponds to 4 phase shifts, namely
$\delta_0=\delta_1=\delta_2=2 \pi$ and $\delta_3=\pi$.
The phase equivalent potential reads
\bea
K_0,r_1 & = &6.80886924,  0.659934689  \nonumber\\
K_1,r_2 & = &1.2151, 12.8731444 \nonumber\\
K_2, r_3 & = &1.34685, 13.8434964  \nonumber\\
K_3, R& = & 1.15320976, 15     \ .
\eea
The values of $n_{\ell}$ are
and $n_1=1,n_2=5,n_3=4.$

The second case includes 5 phase shifts, namely
$\delta_0=\delta_1=\delta_2=\pi$ and $\delta_3=\delta_4=0$.
The phase equivalent potential is given by
\bea
K_0,r_1 & = &  7.43592354,0.604283917 \nonumber\\
K_1,r_2 & = & 1.08584,  5.38409355 \nonumber\\
K_2, r_3 & = & 1.13832,12.2708388   \nonumber\\
K_3, r_4 & = & 1.07046 , 13.398 \nonumber\\
K_4,R & = &  0.814872, 21 \ .
\eea
The values of $n_{\ell}$ are 
$n_1=1,n_2=2,n_3= n_4=3.$

\vspace*{1.cm}

A similar bound is valid for non transparent potentials.
We remind the reader that in this case $n \pi$ 
is the difference between the exact
phase of the equivalent potential and the exact phase of the reference
potential. 
For non transparent potential the above inequality becomes
\be
  n_{\rm min}    
   > \frac{1}{\pi} \left(\pi -c_N \pi-\int_{rt(N,k)}^{R}  \ 
  \sqrt{k^2-v^{(s)}(r')-\frac{N (N+1)}{r'^2} } dr' \right) \ .
\ee
Here, $v^{(s)}$ is the starting reduced  potential and $rt(N,k)$ 
the corresponding largest turning point. 
On the basis of the results obtained  for transparent potentials, we have
\be
n_{\rm min} \geq 1+\left[-\frac{1}{\pi} \left(
\int_{r_1(N)/k}^{R}  \ \sqrt{k^2-\frac{N (N+1)}{r'^2} } dr' +
\delta(N,k) \right) \right] \ .
\label{nminc}
\ee
The quantity
$\delta(N,k)$ is either the exact phase of the starting potential calculated 
from Eq. (\ref{cal}) or the JWKB approximated phase Eq. (\ref{phase}).

For $N=3$ 
and the starting  potential of Table  7 we obtain   
$ n_{\rm min} \geq 1+[-4.08774]=-4$ and $ n_{\rm min} \geq 1+[-4.01251]=-4$ 
 according to the fact that the phase is exact or approximated. 
It has to be compared to the value $n_{\rm min}=-3$ of Table  7.

Let us make a brief comment about the maximal number of nodes $\tilde N$ 
of the wave functions (except the trivial zero at the origin ) inside the potential.  For a given $\ell$, $\tilde N$ is the number of zeros $r_{\ell}$
 of the regular solution of the Schr\"odinger equation below $R$.
   We have checked, from examples of Tables  5-7 and Appendix {\bf B},
that for  all potentials in the "class" $n$ (such that $ (\delta(\ell,k)=n \pi, \ell=0,1,2.3$) 
  $\tilde N$ depends  only on $\ell,k,R,n$.
Moreover $\tilde N(\ell,k,R,n)$ satisfies
 $\tilde N(\ell,k,R,n+1)= \tilde N(\ell,k,R,n)+1$.  
  In this sense, $n$ is correlated to $\tilde N$ as expected from 
 the results of Sabatier \cite{Sabatier}, who
argued that additional multiples of $\pi$ were associated with additional nodes
"inside" the potential.
However the Sabatier's method, based on the Abel transform,
cannot be directly applied here.  
Indeed the latter method  requires the function $r \mapsto r^2  (k^2-v(r))$ 
to be continously differentiable \cite{Sabatier,Cuer}.

\section{Conclusions}
The present work is devoted to the study of discrete ambiguities. They
arise when a specific parametrized expression is used as a potential,
its parameters being fitted  to a finite number of phase-shifts, or
more directly to a scattering amplitude. To this aim, use is made of
piecewise constant potentials, the intervals of which are defined by
the zeros of the regular solution of the Schr\"odinger equation. These
potentials generate a class of phase equivalent potentials with a
phase ambiguity of $n \pi$, $n$ being an integer. The reference
potential belongs obviously to this class.

A few examples have been investigated and solved exactly for 1-5 phase-shifts. The number
of discrete ambiguities depends on the range of the potential. For a
given value of $n$, the phase equivalent potentials can be ordered
according to the position of the zeros of the regular solutions,
except for the zero potential.

Note that the present study  considers both positive and negative
potentials. We have verified that in the case of attractive potentials
having at least one bound state, the measurement of the ground state
energy is a sufficient criterion to fix the potential and get a unique
answer.

For a number of phase-shifts larger than 5, the numerical effort
becomes rapidly intractable for an exact solution. In this case, use
can be made of the JWKB approximation. This last allows to derive a
lower bound for the minimal value $n_{\rm min}$.

Advantage has been taken at several places of the transparent
potentials. They are defined as potentials having all their phase-shifts
equal to  $n \pi$. Particular attention has been given to
the transparent potentials for which $n$ is independent of $\ell$.
Cases can occur, however, such that the exact phases are not the same
for all waves.  Then $n$  depends on $\ell$. 

Moreover, for transparent potentials, a conjecture has been proposed
to determine $n_{\rm min}$.

Finally,  we show that removing  the phase-ambiguity 
 as suggested by Drisko,Satchler and Bassel is not sufficient to
 determine the potential uniquely.
 Potentials reproducing the exact phases
  can differ by the  numbering
 of the zeros of their regular solutions. Actually, the difference
 between the two approaches lies in the following. If a scattering
 amplitude is reproduced with a Woods-Saxon potential, for instance,
 the number of parameters is limited, and the large number of
 phase-shift to be fitted ensure the quasi unique determination of the
 parameters. Thus,  phase ambiguities can be observed, and
 possibly removed. On the other hand, piecewise constant potentials
 generate a much large  functional space. Consequently, ambiguities
 arise also from the choice of the intervals. This behavior has been
 emphasised by Lombard and Wilkin \cite{lw}, in analysing high energy
 differential cross sections via the Glauber model. In this case, there
 is no phase ambiguity but the potential is defined as the statistical
 average over a large ensemble of different piecewise constant 
 potentials fitting the data.

\newpage
{\bf \Large Appendix A}

In this Appendix we report the expression of  $\mathcal{N}_3, \mathcal{D}_3$ of Eq.(\ref{T33})
and  $\mathcal{N'}_3, \mathcal{D'}_3$ of Eq.(\ref{T22n})
\bea
\mathcal{N}_3 & = & 675  D(R) R + 225 R^2 - 90  D(R) K_3^2 R^3 - 15 K_3^2 R^4 - 675  D(R) r_3 \nonumber\\
           &- &225 R r_3 + 315  D(R) K_3^2 R^2 r_3 + 90 K_3^2 R^3 r_3 - 15  D(R) K_3^4 R^4 r_3 \nonumber\\
          & -&   270  D(R) K_3^2 R r_3^2 - 90 K_3^2 R^2 r_3^2 + 36  D(R) K_3^4 R^3 r_3^2 \nonumber\\
          & + &   6 K_3^4 R^4 r_3^2 + 45  D(R) K_3^2 r_3^3 + 15 K_3^2 R r_3^3 \nonumber\\
         &- &21  D(R) K_3^4 R^2 r_3^3 -  6 K_3^4 R^3 r_3^3 +  D(R) K_3^6 R^4 r_3^3 \nonumber\\
\mathcal{D}_3 & = & 675 D(R) + 225 R - 315 D(R) K_3^2 R^2 -  90 K_3^2 R^3 + 15 D(R) K_3^4 R^4 \nonumber\\
          &+& 675 D(R) K_3^2 R r_3 + 225 K_3^2 R^2 r_3 - 90 D(R) K_3^4 R^3 r_3 - 15 K_3^4 R^4 r_3 \nonumber\\
          &- &270 D(R) K_3^2 r_3^2 - 90 K_3^2 R r_3^2 +  126 D(R) K_3^4 R^2 r_3^2 \nonumber\\
       &+ &36 K_3^4 R^3 r_3^2 - 6 D(R) K_3^6 R^4 r_3^2 -  45 D(R) K_3^4 R r_3^3 \nonumber\\
      & -& 15 K_3^4 R^2 r_3^3 + 6 D(R) K_3^6 R^3 r_3^3 + K_3^6 R^4 r_3^3
\label{T33p}
\eea

\bea
{\mathcal{N'}_3} & = & 9 R^2 r_3 - 9 R r_3^2 + 3 K_3^2 R^3 r_3^2 - 3 K_3^2 R^2 r_3^3   +  18 R r_3 C(R)  - 3 K_3^2 R^3 r_3 C(R) \nonumber\\
        & - &18 r_3^2 C(R)  +  9 K_3^2 R^2 r_3^2 C(R) - 6 K_3^2 R r_3^3 C(R) + K_3^4 R^3 r_3^3 C(R)    + 18 R^2 C(r_3)  \nonumber\\
       & - &     18 R r_3 C(r_3) + 6 K_3^2 R^3 r_3 C(r_3) - 9 K_3^2 R^2 r_3^2 C(r_3) + 3 K_3^2 R r_3^3 C(r_3) \nonumber\\
      &- &K_3^4 R^3 r_3^3 C(r_3) + 36 R C(R) C(r_3) - 6 K_3^2 R^3 C(R) C(r_3) - 36 r_3 C(R) C(r_3) \nonumber\\
     & + & 18 K_3^2 R^2 r_3 C(R) C(r_3)-  18 K_3^2 R r_3^2 C(R) C(r_3)  +  3 K_3^4 R^3 r_3^2 C(R) C(r_3) \nonumber\\
     &+ & 6 K_3^2 r_3^3 C(R) C(r_3) -  3 K_3^4 R^2 r_3^3 C(R) C(r_3) \nonumber\\
   {\mathcal{D'}_3} & = & 9 R r_3 - 3 K_3^2 R^3 r_3 + 9 K_3^2 R^2 r_3^2 - 3 K_3^2 R r_3^3  +   K_3^4 R^3 r_3^3 + 18 r_3 C(R)  \nonumber\\
     &- & 9 K_3^2 R^2 r_3 C(R) + 18 K_3^2 R r_3^2 C(R) -  3 K_3^4 R^3 r_3^2 C(R) - 6 K_3^2 r_3^3 C(R)  \nonumber\\
      &+ & 3 K_3^4 R^2 r_3^3 C(R)  + 18 R C(r_3) - 6 K_3^2 R^3 C(r_3) + 18 K_3^2 R^2 r_3 C(r_3) - 9 K_3^2 R r_3^2 C(r_3) \nonumber\\
     &+ &3 K_3^4 R^3 r_3^2 C(r_3) - 3 K_3^4 R^2 r_3^3 C(r_3) + 36 C(R) C(r_3) - 18 K_3^2 R^2 C(R) C(r_3) \nonumber\\
     &+ & 36 K_3^2 R r_3 C(R) C(r_3) - 6 K_3^4 R^3 r_3 C(R) C(r_3)  -  18 K_3^2 r_3^2 C(R) C(r_3) \nonumber\\
     &+ &9 K_3^4 R^2 r_3^2 C(R) C(r_3) -  6 K_3^4 R r_3^3 C(R) C(r_3) + K_3^6 R^3 r_3^3 C(R) C(r_3)
\eea

\newpage
\begin{table}
{\bf \Large Appendix B}
\vspace*{.5 cm}
\begin{center}
\begin{tabular}{|r|r|r|r|r|r|r|r|r|r|r|r|}
\hline
$ n$ & $n_1$ & $n_2$ & $n_3$ & $K_0$ & $r_1$ & $ K_1$ & $r_2$&  $K_2$ & $r_3$ & $K_3$ & $R$   \\
\hline
  1 & 1 & 1 & 1 &   1.737&   2.587&   3.446&   2.990&   2.883&   3.118&   2.456&  10.000 \\
   1&1 & 1 & 2 &   0.918&   4.896&   3.900&   5.275&   1.855&   7.097&   4.680&  10.000 \\
    1& 1 & 1 & 3 &   1.932&   2.325&   3.475&   2.722&   2.673&   5.356&   2.209&  10.000 \\
  1 & 1 & 1 & 4 &    1.996&   2.252&   3.437&   2.650&   2.593&   6.627&   2.075&10.000 \\ 
   1& 1 & 1 & 5 &    2.057&   2.184&   3.361&   2.589&   2.499&   8.024&   1.895&  10.000 \\
    1 & 1 & 2 & 2 &    1.943&   2.312&   3.476&   3.635&   2.737&   3.805&   2.197&  10.000 \\
  1& 1 & 2 & 3 &   2.106&   2.134&   3.515&   3.442&   2.596&   4.903&   2.019&  10.000 \\
  1& 1 & 2 & 4 &    2.235&   2.011&   3.505&   3.309&   2.473&   6.180&   1.832& 10.000 \\
     1 & 1 & 2 & 5 &   2.360&   1.904&   3.456&   3.233&   2.339&   7.675&   1.595&  10.000   \\
   1& 1 & 3 & 3 &    2.171&   2.070&   3.492&   4.302&   2.549&   4.518&   1.886&  10.000  \\
1& 1 & 3 & 4 &  2.388&   1.882&   3.539&   4.085&   2.373&   5.716&   1.639&10.000 \\
   1 & 1 & 3 & 5 &    2.592&   1.734&   3.539&   3.938&   2.187&   7.225&   1.332&  10.000 \\
   1& 1 & 4 & 4 &   2.395&   1.877&   3.475&   5.035&   2.309&   5.309&   1.506&10.000 \\
  1& 1 & 4 & 5 &    2.648&   1.697&   3.520&   4.818&   2.068&   6.729&   1.139&  10.000 \\
 1& 1 & 5 & 5 &   2.514&   1.787&   3.367&   5.991&   2.009&   6.339&   1.032&  10.000 \\
 1 & 2 & 2 & 2 &    2.032&   3.802&   4.212&   4.148&   2.772&   4.323&   2.384&  10.000 \\
  1 & 2 & 2 & 3 &    1.324&   5.834&   5.330&   6.117&   2.084&   7.739&   4.610&  10.000 \\
  1 & 2 & 2 & 4 &     2.228&   3.467&   4.100&   3.821&   2.573&   6.558&   2.032&  10.000 \\
   1 & 2 & 2 & 5 &    2.327&   3.320&   3.974&   3.682&   2.441&   7.909&   1.783&  10.000 \\
    1& 2 & 3 & 3 &    2.208&   3.498&   4.069&   4.634&   2.589&   4.852&   2.000&  10.000 \\
  1& 2 & 3 & 4 &    2.378&   3.248&   4.027&   4.395&   2.428&   5.986&   1.745&  10.000 \\
    1& 2 & 3 & 5 &    2.544&   3.037&   3.923&   4.189&   2.238&   7.393&   1.418&  10.000 \\
   1& 2 & 4 & 4 &    2.389&   3.234&   3.912&   5.225&   2.336&   5.500&   1.565&  10.000 \\
  1& 2 & 4 & 5 &   2.615&   2.954&   3.867&   4.967&   2.094&   6.856&   1.182&  10.000 \\
    1& 2 & 5 & 5 &    2.483&   3.112&   3.691&   6.078&   2.019&   6.429&   1.054&  10.000 \\
    1 & 3 & 3 & 3 &   2.346&   4.647&   4.825&   4.956&   2.639&   5.171&   2.126&  10.000 \\
    1 & 3 & 3 & 4 &    2.480&   4.398&   4.691&   4.714&   2.492&   6.258&   1.869&  10.000 \\
  1& 3 & 3 & 5 &   2.628&   4.150&   4.490&   4.479&   2.303&   7.581&   1.530&  10.000 \\
  1 &   3 & 4 & 4 &   2.488&   4.382&   4.479&   5.418&   2.369&   5.692&   1.631&  10.000 \\
   1& 3 & 4 & 5 &    2.689&   4.056&   4.326&   5.127&   2.125&   6.989&   1.231&  10.000 \\
 1& 3 & 5 & 5 &    2.551&   4.274&   4.111&   6.170&   2.033&   6.524&   1.079&  10.000 \\
  1 & 4 & 4 & 4&   2.657&   5.293&   5.134&   5.586&   2.401&   5.858&   1.693&  10.000 \\
 1 & 4 & 4& 5 &    2.833&   4.966&   4.864&   5.273&   2.155&   7.109&   1.280&  10.000 \\
  1 & 4& 5 & 5 &   2.685&   5.240&   4.592&   6.252&   2.046&   6.608&   1.103&  10.000 \\
1 & 5 & 5 & 5 &    2.858&   6.026&   5.112&   6.322&   2.059&   6.678&   1.124&  10.000 \\
\hline
\end{tabular}
\end{center}
\caption{ This Table is an extension of Table 7 (see text).}
\end{table}

\end{document}